\documentclass[a4paper,12pt]{article}
\usepackage{graphicx}
\usepackage{times}
\usepackage[dvips,unicode,colorlinks,linkcolor=blue,citecolor=blue,urlcolor=blue]{hyperref} %\textwidth=14.5 cm
\def\ddot{{\displaystyle{{{\partial^2}\over{\partial t^2}}}}}

\begin{document}

%\maketitle
\title{Discrete optic breathers in zigzag
chain: analytical study and computer
simulation}
\maketitle

\vskip1pt \centerline{L. I. Manevitch,~~~  A. V.
Savin} \vskip1pt
 \centerline{Semenov Institute of Chemical Physics}
 \vskip1pt
\centerline{Russian Academy of Sciences,
      Moscow, 117977, Russia}
      \vskip1pt
\centerline{lmanev@center.chph.ras.ru $\quad$
     asavin@center.chph.ras.ru}
\vskip10pt
\centerline{and}
\vskip10pt
\centerline{C.-H. Lamarque}
\vskip1pt
\centerline{ENTPE/DGCB/LGM}
\vskip1pt
\centerline{3, rue Maurice Audin,F69518 Vaulx-en-Velin Cedex France}
\vskip1pt
\centerline{lamarque@entpe.fr}

\tableofcontents
\newpage
\begin{abstract}
We present analytical and numerical
study of discrete breathers identified as localized deformations
of valence angles accompanied by change of valence bonds in
crystalline polyethylene (PE). It is shown that  such breathers
can exist inside the optic wave number band and can propagate along the
macromolecule chain with subsonic velocities. Analytical results
are confirmed by numerical simulation using Molecular Dynamics
procedure. We examine  also stability of the breathers relative to
thermal excitations and mutual collisions as well as to collisions
with acoustic non-topological and topological solitons. The
conditions of breathers existence depend strongly on their
generating frequency, relationship between stiffnesses of valence
bonds and valence angles  as well as on type of nonlinear
characteristics of intrachain interactions.
\end{abstract}
\vskip1pt

%\keywords
%{Keywords:}
%{Nonlinear normal modes, hamiltonian, infinite
%chain, branch, breathers, acoustic solitons, topological solitons}

\section{Introduction}
In spite of growing interest to study of localized nonlinear oscillatory
excitations, almost all studies in this field are devoted to straight chains.
The reason is that even the most simple realistic models of strongly
anisotropic systems, e.g. polymer crystals, turn out to be substantially more
complicated systems than commonly considered one-dimensional models.

\noindent
Two-dimensional linear dynamics of planar zigzag chain was considered more than
60 years ago by Kirkwood \cite{kirkwood1939} in application to polyethylene
macromolecule. From the other side zigzag and helix chains can be also
considered as the discrete models of corrugated and spiral mechanical systems
\cite{Magno2002}. Growing interest to nonlinear dynamics of polymer chains and
crystals in last decades is stimulated by numerous applications to such
physical problems as dielectric relaxation \cite{Skinner80,Boyd85,Zubova2002},
premelting and melting \cite{Ginzburg91}, heat conductivity
\cite{Gendelman92,Gendelman2000}, polarization \cite{Taylor80}, fracture
\cite{Manevitch2001,Manevitch90}, plastic deformation
\cite{Ginzburg92a,Ginzburg92b,p14,p15}, reading the information in DNA
\cite{Yakushevich2002}.  All these applications are based on two types of
solitons : supersonic solitons of tension and compression
\cite{Manevitch2001,Manevitch97,p18} and combined soliton of tension
(compression) and torsion \cite{Manevitch2001,p19,p20,p21,p22}, both types
being in long wavelength region.

\noindent
Besides there are two examples of breathers in the models of polymer chains.

\noindent First one is localized nonlinear vibration of C--H
valence bond in carbon-hydrogen chains due to anharmonicity of
corresponding potential \cite{p23}. Such a breather has rather
high frequency ($\simeq 3100$ cm$^{-1}$) because it is higher than
lower optic branch of IR spectrum. Second example is localized
periodic change of valence angles in PE chains accompanied by
change of valence bonds \cite{Manevitch2003}. This breather
revealed by numerical simulation has significantly lower
fundamental frequency ($\simeq 800$ cm$^{-1}$) that is close to
lower boundary of first optic zone for PE crystals. However, one
type of motion (namely transversal one) turned out to be
predominant in the localization region of this breather because
its spatial frequency is close to boundary value of wave number
($k =\pi$). Besides, only stationary breathers have been revealed.
Meanwhile, the valences angle potential for PE chain presented in
\cite{p20} demonstrates the case when lower boundary of optic
dispersion curve is inside of wave number diapason and is not
close to its boundaries. It means that possible breather in this
case has not a predominant component (longitudinal or transversal)
and its study becomes much more complicated.

In this paper we remove both restrictions mentioned above.
Besides, we present first analytical description of short
wavelength nonlinear dynamics of PE chain in all diapason of optic
frequencies alongside with numerical investigation of breathers
stability relative to thermal excitations as well as mutual
collisions and collisions ith acoustic non-topological and
topological solitons. It worth mentioning that optic breathers in
PE crystal weakly depend on interchain interaction (contrary to
topological solitons). However, this interaction was taken into
account in the course  of numerical studies.

\noindent The commonly studied breathers in attenuation band can
be, in principle, linearly stable \cite{p25}, \cite{p27}. However
as it was shown in \cite{Manevitch2003}, their  nonlinear
interaction with extended modes (phonons) leads to finite times of life in
realistic models of polymer crystals for breathers. Therefore, in
such models there is not qualitative difference between the
breathers in attenuation and propagation bands, if naturally they
can exist in propagation band at all. From other side, the finite
time of life does not  mean that such breathers are not of physical
importance. They can be manifested in different physical
properties \cite{p26} and their instability relative to
interaction with phonons leads to non-monotonous dependence of
breathers contribution with increasing the temperature
\cite{Manevitch2003}.
%---------------------------- Fig. 1 ------------------------------------
\begin{figure}[tb]
\begin{center}
\includegraphics[angle=0, width=1\linewidth]{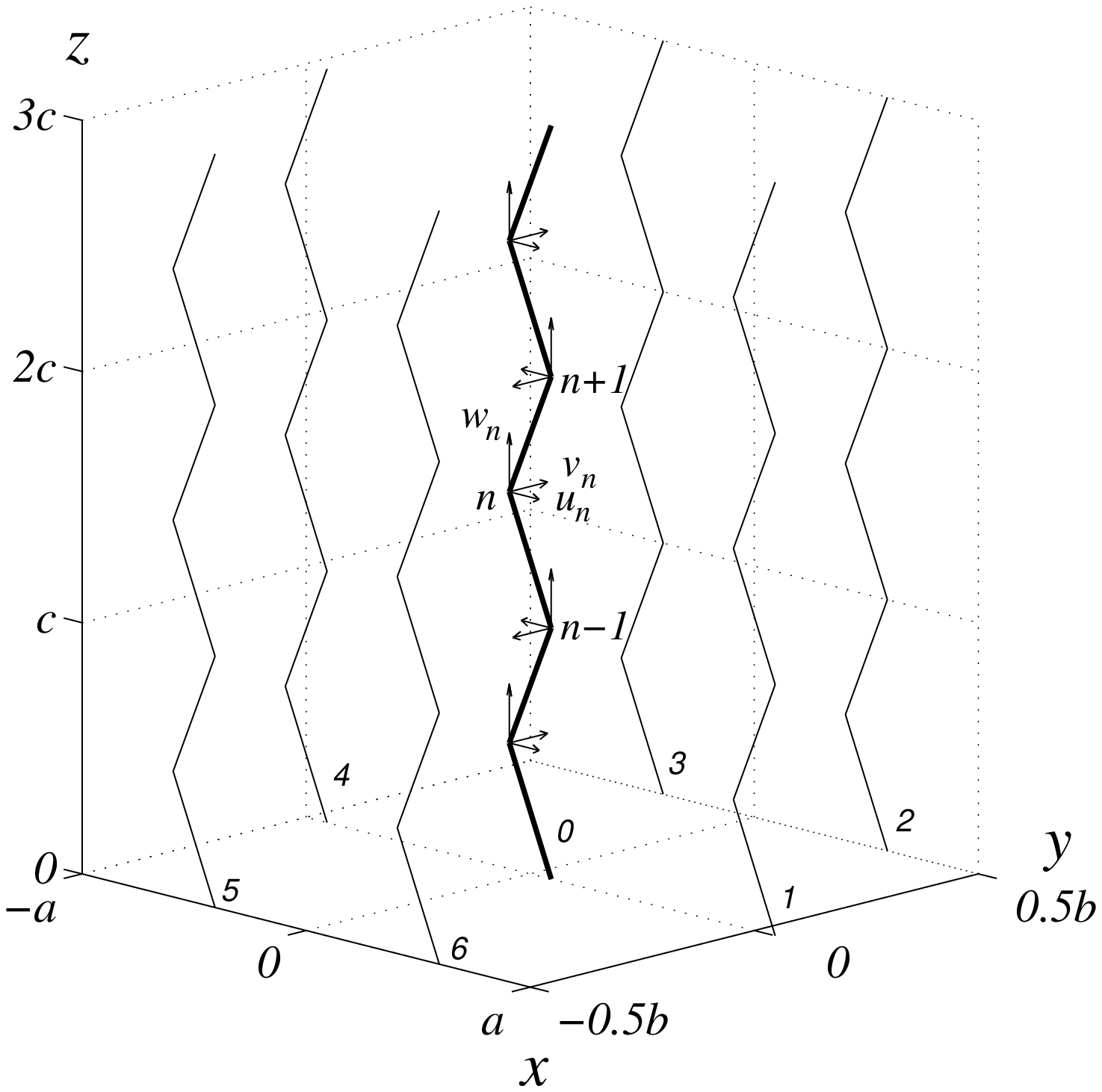}
\end{center}
\caption{\label{fg01}\protect
Schematic presentation of monoclinic structure of crystalline PE.
The considered chain (curve 0) with local coordinates and 6 neighbor chains
(curves 1-6) are shown.
}
\end{figure}
%---------------------------- Fig. 1 ------------------------------------

\section{Description of the model}
Because in numerical study we deal not only with planar motion of
the chain but take also into account the spatial movement and
interchain interactions, three-dimensional hamiltonian is
presented below.

\noindent
It is suggested that the zig-zag initial structure of PE macromolecule is
directed along the $z$-axis in plane ($x,z$), while $y$-axis is perpendicular
to the plane $(x,z)$ so that $(x,y,z)$ is the positive global reference of the
problem, $\theta_0$ being the angle of the zigzag chain.

\noindent
Initial coordinates of the $n$-th mass of the chain are :
\begin{equation}
z_n^0 = n l_z,~~~ x_n^0 = (-1)^{n+1} l_x /2, \label{f1}
\end{equation}
where $l_z=s\rho_0$ and $l_x=c\rho_0$ [$s = \sin(\theta_0/2)$, $c =
\cos(\theta_0/2)$) are longitudinal and transversal dimensions of the initial
zigzag chain. We use approximation of "united atoms" because relative motions
of hydrogen atoms are non-significant when dealing with backbone deformation
predominantly. So the magnitude of every mass is supposed to be equal to 14
a.u. It is convenient to introduce the relative coordinates
\begin{equation}
u_n = x_n - x_n^0,~~~ w_n = z_n - z_n^0, \label{f2}
\end{equation}
where $w_n$, $u_n$, $v_n$ determine the longitudinal, transversal and
out-of-plane displacements of the $n$-th mass from its equilibrium state. Let
us denote by $\rho_n$ and $\theta_n$ the current length of the valence bond and
current valence angle respectively. We introduce also $\phi_n$ as the angle
between $d_{n-1}$ and the plane generated by $d_n$  and $d_{n+1}$ (conformation
angle).

\noindent
The dynamics of the chain is governed by the following hamiltonian function of
the chain  \cite{Manevitch2003}:
\begin{eqnarray}
H&=&\sum_{n=-\infty}^{+\infty}\{\frac12m(\dot u_n^2+\dot v_n^2+\dot w_n^2)
 \nonumber \\
&& +[p_1 + p_2 \cos(\phi_n) + p_3\cos(3\phi_n)] \nonumber \\
&& +\frac12\Lambda [\cos(\theta_n)-\cos(\theta_0)]^2 \nonumber \\
&& +{\cal D}[1 - e^{-q(\rho_n - \rho_0)}]^2 \} + Z(u_n,v_n,w_n)
\label{f3}
\end{eqnarray}
where dots denote the time derivation, $\Lambda$ and ${\cal D}$
correspond to energies of valence angles and valence bonds
respectively, and the last term the energy of interaction of the
n$^{th}$ unit with six neighboring chains (substrate potential).
Parameters $p_1, p_2, p_3$ satisfy conditions $p_1 = p_2 +p_3$,
$p_2,p_3 > 0$, so that undeformed state corresponds to $\phi_n =
\pi$. Let us set $k_1 = 2 {\cal D} q^2$ and $k_2 = \Lambda
\sin^2\theta_0$. Then
\begin{equation}
\rho_n^2 = (u_{n+1}-u_n -l_x)^2 + (v_{n+1}-v_n)^2 + (w_{n+1}-w_n + l_z)^2.
\label{f4}
\end{equation}
We have also:
\begin{eqnarray}
\cos\theta_n&=&(-a_{n-1,1}a_{n,1}-a_{n-1,2}a_{n,2}\nonumber\\
            && -a_{n-1,3}a_{n,3})/\rho_n\rho_{n-1}, \label{f5} \\
\cos\phi_n&=&(b_{n,1}b_{n+1,1}+b_{n,2}b_{n+1,2}\nonumber\\
            && +b_{n,3}b_{n+1,3})/\beta_n\beta_{n+1},
\label{f6}
\end{eqnarray}
where
\begin{eqnarray*}
b_{n,1}&=&a_{n-1,2}a_{n,3}-a_{n,2}a_{n-1,3}, \\
b_{n,2}&=&a_{n-1,3}a_{n,1}-a_{n,3}a_{n-1,1}, \\
b_{n,3}&=&a_{n-1,1}a_{n,2}-a_{n,1}a_{n-1,2}
\end{eqnarray*}
are related to inner product vector coordinates,
and $\beta_{n}=(b_{n,1}^2 + b_{n,2}^2 + b_{n,3}^2)^{1/2}$
are norms of corresponding vectors with
\begin{eqnarray*}
a_{n,1} &=& u_{n+1} - u_n + l_x,\\
a_{n,2} &=& v_{n+1} - v_n,\\
a_{n,3} &=& w_{n+1}-w_n + l_z, \\
\rho_n&=&(a_{n,1}^2 + a_{n,2}^2 + a_{n,3}^2)^{1/2},
\end{eqnarray*}
which are the vector coordinates. The substrate potential
$$
Z(u,v,w) = \varepsilon_w \sin^2(\pi w/l_z) +{1 \over 2} K_u [1 +
\varepsilon_u \sin^2({{\pi w}\over{ l_z}})] \{ u - {1 \over 2} l_x
[1 - \cos(({{\pi w}\over{ l_z}}) ]\}^2 + {1 \over 2} K_v [1 +
\varepsilon_v \sin^2({{\pi w}\over{ l_z}})]  v^2,
$$ where
$\varepsilon_u = 0.067 426 5$ kJ/mol, $\varepsilon_v = 0.041 835
3$ kJ/mol, $\varepsilon_w = 0.149 012 4$ kJ/mol, $K_u= 2.169 513$
kJ/ $\AA$ mol$^2$, $K_v= 13.683 865$ kJ/ $\AA$ mol$^2$.

\section{Planar motion of zigzag chain}
We consider further analytically only in-plane dynamics of zigzag. In both
linear and nonlinear problems such a dynamics can be fully separated from
out-of-plane motion. Therefore, general Hamiltonian can be simplified and one
can obtain in both physically and geometrically nonlinear approach:
\begin{eqnarray}
 H &=& \sum_{n=-\infty}^{+\infty}\frac12m(\dot u_n^2+\dot w_n^2)
 \mbox{ terms up to order 4 from} \nonumber \\
&&\Biggl\{\sum_{n=-\infty}^{+\infty}\frac12\Lambda[\cos(\theta_n)-\cos(\theta_0)]^2
\nonumber \\
&&+ \sum_{n=-\infty}^{+\infty}{\cal D}  [1 - e^{-q(\rho_n - \rho_0)}]^2
 \Biggr\}
\label{f7}
\end{eqnarray}
There are two systems of parameters describing PE macromolecules
which are supposed in \cite{p19} and \cite{p20} respectively which
differ by value of parameter $k_2$ ($k_2/\sin^2\theta_0=130.1$
kJ/mol in \cite{p19} and 529 kJ/mol in \cite{p19}). We will
consider both systems supposing \cite{p18,p19} that
$k_1/q^2=334.7$ kJ/mol, $q=19.1$ nm$^{-1}$,
$\theta_0=113^{\circ}$, $\rho_0 =1.54$~\AA $\,$ so that
$\delta=k_2/k_1\rho_0^2=0.019$ \cite{p19} or 0.078 \cite{p20}.

\noindent
The deformations are presented by their power expansions including
the terms of first and second order with respect to displacements:
\begin{eqnarray}
\Delta\rho_n&=&\rho_n-\rho_0 =\nonumber\\
&\quad &(u_n -u_{n+1})c+(w_{n+1}-w_n)s+ {1 \over {2 \rho_0}} ((u_n
-u_{n+1})c+(w_{n+1}-w_n)s)^2 + \dots,
\label{f8} \\
%\Delta\cos_n&=&\cos\theta_n-\cos\theta_0\nonumber\\
\Delta\theta_n&=&\theta_n-\theta_0 = \nonumber\\ & \quad &
\frac{s}{\rho_0}(2u_n-u_{n-1}-u_{n+1}) + \frac{c}{\rho_0} (w_{n-1}-w_{n+1})\nonumber\\
&&+ \frac{cs}{\rho_0^2}[(u_n-u_{n+1})^2 +
(u_n-u_{n-1})^2 - (w_n-w_{n+1})^2 - (w_n-w_{n-1})^2] \nonumber\\
&&+ \frac{c^2- s^2}{\rho_0^2}[(w_n-w_{n-1})(u_n-u_{n-1}) +
(w_{n+1}-w_n)(u_n-u_{n+1})]+... \label{f9}
%\frac{\sin\theta_0}{\rho_0}[(2u_n-u_{n-1}-u_{n+1})s\nonumber\\
%&&+(w_{n-1}-w_{n+1})c]+ \mbox{ h.o.} \label{f9}
\end{eqnarray}
\noindent
Equations of motion are obtain in the form:
\begin{equation}
m\ddot{u}_n=-\frac{\partial H}{\partial u_n},~~~
m\ddot{w}_n=-\frac{\partial H}{\partial w_n} \label{hamn}
\end{equation}
for the $n$th particle.
\vskip10pt\noindent
Linearized equations
are given by
\begin{eqnarray}
m\ddot u_n &+& k_1 sc(w_{n+1}-w_{n-1})
+ k_1c^2 (2 u_n-u_{n-1}-u_{n+1})\nonumber\\
&+&k_2cs(w_{n-2}+2w_{n-1}-2w_{n+1}-w_{n+2})/\rho_0^2 \nonumber\\
&+&k_2s^2(u_{n-2}-4 u_{n-1}+6u_n -4 u_{n+1}+u_{n+2})/\rho_0^2 = 0
%\mbox{ h.o.t.}
\label{f10}\\
m\ddot w_n&-&k_1s^2(w_{n+1}-2w_n+w_{n-1})
+k_1sc(u_{n+1}-u_{n-1})\nonumber\\
&-&k_2c^2(w_{n-2}-2w_{n}+w_{n+2})/\rho_0^2 \nonumber\\
&+& k_2s^2(-u_{n-2}+2
u_{n-1}-2u_{n+1} +u_{n+2})/\rho_0^2 = 0
%\hbox{ h.o.t. }
\label{f11}
\end{eqnarray}

\section{Dispersion relations}
The linear dispersion curves have well known form (see, e.g.
\cite{Manevitch2003}),
which is described by relation :
\begin{equation}
\omega^2(k) = f(k) = \omega_0^2(k) \pm \sqrt{\omega_0^4(k) -
\omega_1^4(k)}
%\label{eq9}
\label{f12}
\end{equation}
where
\begin{eqnarray}
\omega_0^2(k) &=& C_1 (1 - \cos\theta_0\cos k)\nonumber\\
&&+2C_2(1-\cos k)(1+\cos k\cos\theta_0),
\label{f13}\\
\omega_1^4(k)&=&8C_1C_2(1- \cos k)\sin^2k.
%\label{f14}
\nonumber
\end{eqnarray}
Here, $C_1 =k_1/m$, $C_2=k_2/m\rho_0^2$, $k=\tilde{k}l_z$, $\tilde{k}$ is wave number. Signs
"minus" and "plus" correspond to acoustic and optic branches respectively. We
consider further the optic branch only.
%
%Because existence of breathers depends strongly on curvature of
%dispersion curve, we will deal with three characteristic cases:
%$k\approx 0,~\pi,~\pi/2$.
%
Corresponding asymptotic representations of the linear frequencies
%in the vicinities of these wave numbers
can be presented as follows in the vicinity of arbitrary wave number
$k^\star$:
\begin{equation}
 \omega^2(k) =
 %f(k) =
 \Omega^2 + \nu (k-k^\star) + \mu
(k-k^\star)^2 + \dots = f(k), \label{f15}
\end{equation}
with
%\begin{eqnarray}
%\Omega^2 &=& f(k^\star), \nonumber\\
%\nu &=&  f'(k^\star)=,\label{f16}\\
%\mu &=& f"(k^\star)= , \nonumber
%\end{eqnarray}
\begin{eqnarray}
\Omega^2 &=& 2C_1[1-\cos(\theta_0)\cos(k^\star)]- {{4C_2\cos^2(k^\star)\sin^2(\theta_0)}\over{1-\cos(\theta_0)\cos(k^\star)}} [1-\cos(k^\star)],
 \nonumber\\
\nu &=&  2C_1\sin(k^\star)\cos(\theta_0) \nonumber\\ &&+
{{2C_2\sin(2k^\star)\sin^2(\theta_0)}\over{(1-\cos(\theta_0)\cos(k^\star))^2}}
[(1-\cos(k^\star))(2\cos(k^\star)-1) + \cos(\theta_0)\cos(k^\star) -1],\label{f16}\\
\mu &=& C_1\cos(\theta_0)\cos(k^\star)+{{C_2
\sin^2(\theta_0)}\over{(\cos(\theta_0)\cos(k^\star)-1)^3}}[-4+12\cos(k^\star)
\nonumber \\&&+4\cos^2(k^\star)(2-3\cos(\theta_0))
-2\cos^3(k^\star) (-2\cos^2(\theta_0)+3\cos(\theta_0)+9) \nonumber
\\&&
+2\cos^4(k^\star)\cos(\theta_0)(\cos(\theta_0)+11)-8\cos^5(k^\star)\cos^2(\theta_0)]
. \nonumber
\end{eqnarray}
\noindent
Table~\ref{tabledisp} provides values of $\Omega^2, \nu,\mu$ for some particular $k$.
\begin{table}[tbh]
\caption{Values of $\Omega^2, \nu,\mu$ for some particular $k$.}
\begin{center}
\begin{tabular}{ccccc}
\hline\hline \\
$k$ & $\Omega^2$ & $\nu$ & $\mu$ \\ \\
\hline\hline \\
$0$& $4 s^2 C_1$ & 0 & $C_1(\cos(\theta_0)+4c^2\delta)$\\ \\
\hline \\
$\pi$&  $4C_1(c^2+4s^2\delta)$ & 0 & $-C_1\displaystyle{{{\cos(\theta_0)c^2+20c^2s^2\delta}\over{c^2+4s^2\delta}}}$ \\ \\
\hline \\
$\pi/2$& $2C_1$ & $2C_1\cos(\theta_0)$ & $\displaystyle{4C_1C_2 {{\sin^2(\theta_0)}\over{C1-2C_2}}}$\\
\\ \hline
 \end{tabular}
\end{center}
\label{tabledisp}
\end{table}

\begin{figure}[tb]
\begin{center}
\includegraphics[angle=0, width=1\linewidth]{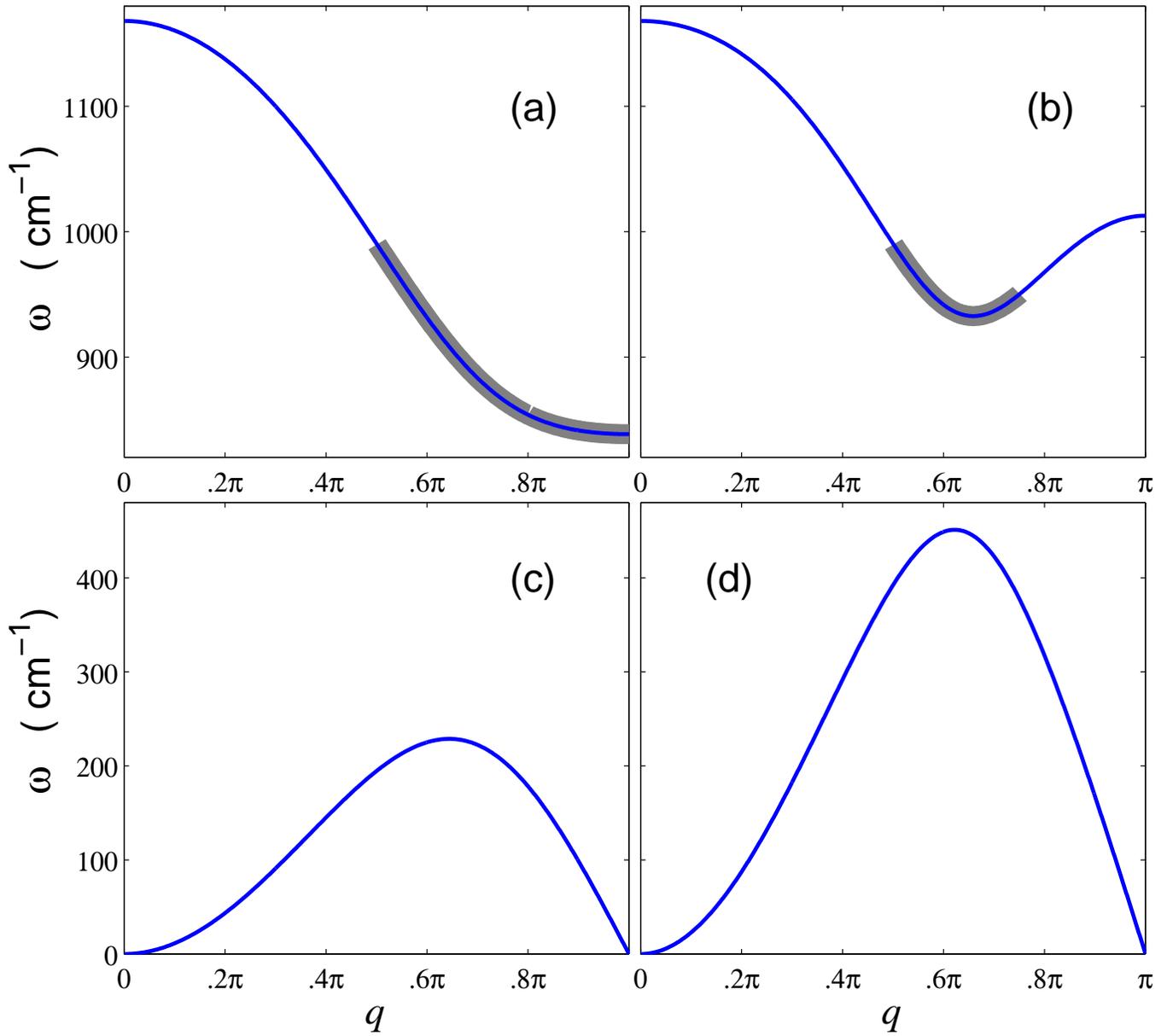}
\end{center}
\caption{\label{fg02}\protect
Dispersion curves: optic (a, b) and acoustic (c, d) branches
for $\delta=0.019$  (a, c) and for $\delta=0.078$ (b, d) (parameter
$C_2=\delta C_1$). Thick part of the dispersion curves correspond
to possible existence of the breathers.
}
\end{figure}

\noindent
Analysis of general expressions of dispersion curves in
application to realistic values of chain parameters reveals two
types of behavior.
 The corresponding plots of dispersion curves are presented in Fig. \ref{fg02}.
% old figure 2!  \ref{disp1a},\ref{disp1o} and \ref{disp2a}, \ref{disp2o}.
% old figure 2!
% old figure 2!
% old figure 2!  \begin{figure}[tb]
% old figure 2!  \begin{center}
% old figure 2!  \includegraphics[angle=0, width=1\linewidth]{acoustic019.ps}
% old figure 2!  \end{center}
% old figure 2!  \caption{\label{disp1a}\protect Dispersion curve: acoustic branch,
% old figure 2!  $\delta=0.019$, $C_2=\delta C_1$. }
% old figure 2!  \end{figure}
% old figure 2!
% old figure 2!
% old figure 2!  \begin{figure}[tb]
% old figure 2!  \begin{center}
% old figure 2!  \includegraphics[angle=0, width=1\linewidth]{optic019.ps}
% old figure 2!  \end{center}
% old figure 2!  \caption{\label{disp1o}\protect Dispersion curve: optic branch,
% old figure 2!  $\delta=0.019$, $C_2=\delta C_1$. }
% old figure 2!  \end{figure}
% old figure 2!
% old figure 2!
% old figure 2!  \begin{figure}[tb]
% old figure 2!  \begin{center}
% old figure 2!  \includegraphics[angle=0, width=1\linewidth]{acoustic078.ps}
% old figure 2!  \end{center}
% old figure 2!  \caption{\label{disp2a}\protect Dispersion curve: acoustic branch,
% old figure 2!  $\delta=0.078$, $C_2=\delta C_1$. }
% old figure 2!  \end{figure}
% old figure 2!
% old figure 2!
% old figure 2!  \begin{figure}[tb]
% old figure 2!  \begin{center}
% old figure 2!  \includegraphics[angle=0, width=1\linewidth]{optic078.ps}
% old figure 2!  \end{center}
% old figure 2!  \caption{\label{disp2o}\protect Dispersion curve: optic branch,
% old figure 2!  $\delta=0.078$, $C_2=\delta C_1$. }
% old figure 2!  \end{figure}
%---------------------------- Fig. 2 ------------------------------------
%---------------------------- Fig. 2 ------------------------------------

\noindent Let us introduce $C_2=\delta C_1$.
The signs of numerical values of optic dispersion curves curvature $\cal{H}$
are:
\begin{itemize}
\item $\delta= 0.019$, ${\cal H}(0) < 0$, ${\cal H}(\pi)> 0$,
${\cal H}(\pi/2)<0$,

\item $\delta= 0.078$, ${\cal H}(0)< 0$, ${\cal H}(\pi)< 0$,
${\cal H}(\pi/2)> 0$.
\end{itemize}
We can see that for $k\approx0$ curvature is negative in both
cases. For $k\approx\pi$, $k\approx\pi/2$ it could be both
positive or negative depending on the magnitude of $\delta$.
\noindent The signs of the curvature determine as we will see the
signs of second spatial derivatives in final continuum equations
of motion. \noindent Let us define $k_0 \in ]0, \pi[$ so that the
corresponding coefficient $\nu$ is equal to zero: $f'(k_0) = 0$
($k_0$ is defined only for $\delta=0.078$, because for
$\delta=0.019$ the value $\nu=0$ at boundaries of wave number diapason
only).

\section{Introduction of modulating functions}

In this section we introduce slow modulating functions in order to
extend analysis based on normal modes that can be done for
linearized equations to the nonlinear case. \noindent So we
consider equations for $u_n, w_n$ both with modulating
continuous space functions $W, {\tilde W}$ expanded close to
arbitrary n$^{th}$ particle: \begin{equation} u_{n+m} =  \cos(mk)
[U + m\epsilon {{\partial U}\over{\partial \xi}}+ {1 \over 2}
m^2\epsilon^2 {{\partial^2 U}\over{\partial \xi^2}}] + \sin(mk)
[{\tilde U} + m\epsilon {{\partial {\tilde U}}\over{\partial
\xi}}+ {1 \over 2} m^2\epsilon^2 {{\partial^2 {\tilde
U}}\over{\partial \xi^2}}]+ \dots \label{modunU}
\end{equation}
\begin{equation} w_{n+m} = \cos(mk) [W + m\epsilon {{\partial W}\over{\partial \xi}}+ {1 \over 2} m^2\epsilon^2 {{\partial^2 W}\over{\partial \xi^2}}]
+ \sin(mk) [{\tilde W} + m\epsilon {{\partial {\tilde
W}}\over{\partial \xi}}+ {1 \over 2} m^2\epsilon^2 {{\partial^2
{\tilde W}}\over{\partial \xi^2}}] + \dots \label{modunW}
\end{equation}
where $\epsilon$ is a small parameter characterizing distance between particles
in the units $\xi=\varepsilon^{-1}z/l_z$, $z$ being dimensional distance.

In the same spirit, we consider equations for $u_{n+1}, w_{n+1}$, both with the modulating
functions expanded close to the ${n+1}^{th}$ particle (i.e. now with
$u_{n+1}= {\tilde U}$, $w_{n+1} = {\tilde W}$):
\begin{equation}
u_{n+1+m} =  \sin(-mk) [U + m\epsilon {{\partial U}\over{\partial \xi}}+ {1 \over 2} m^2\epsilon^2 {{\partial^2 U}\over{\partial \xi^2}}]
+ \cos(mk) [{\tilde U} + m\epsilon {{\partial {\tilde
U}}\over{\partial \xi}}+ {1 \over 2} m^2\epsilon^2 {{\partial^2
{\tilde U}}\over{\partial \xi^2}}]+ \dots \label{modun1U}
\end{equation}
\begin{equation} w_{n+1+m} = \sin(-mk) [W + m\epsilon {{\partial W}\over{\partial \xi}}+ {1 \over 2} m^2\epsilon^2 {{\partial^2 W}\over{\partial \xi^2}}]
+ \cos(mk) [{\tilde W} + m\epsilon {{\partial {\tilde
W}}\over{\partial \xi}}+ {1 \over 2} m^2\epsilon^2 {{\partial^2
{\tilde W}}\over{\partial \xi^2}}] + \dots \label{modun1W}
\end{equation}
\noindent
Necessity to consider the expansions in the vicinities of two points
is caused by twice degeneration of linear normal normal modes
spectrum (except the boundary values of wave number).
By substituting  modulating functions in the linearized equations for
$u_n$, $u_{n+1}$, $w_n$, $w_{n+1}$ (\ref{f10}), (\ref{f11}), one  obtains 4 partial
differential equations in the general case that take into account  linear
part of the equations for modulations of normal modes. Linearized
equations for modulations of normal modes have to be in accordance
with dispersion relations presented above. We present them below
for special case $k = \pi/2$.
%permit us to check dispersion relation. For general case
%expressions are very complicated and are not given here. For
%$k=\pi/2$, these expressions corresponds to
\begin{equation}
\matrix{ \displaystyle{ {{\partial^2 U}\over{\partial t^2}}-2(C_1 c^2 + 4 C_2 s^2) \epsilon {{\partial \tilde{U}}\over{\partial \xi}}
- 4 cs C_2 \epsilon {{\partial W}\over{\partial \xi}} - 4 C_2 s^2
\epsilon^2 {{\partial^2 U}\over{\partial \xi^2}} }\cr \quad \cr
\displaystyle{ + sc(2C_2 + C_1) \epsilon^2 {{\partial^2 {\tilde
W}}\over{\partial \xi^2}} + 2(c^2 C_1 + 2 s^2 C_2) U +
2sc(2C_2-C_1) {\tilde W} = 0,}\cr}
\label{fm1}
\end{equation}
\begin{equation}
\matrix{ \displaystyle{ {{\partial^2 W}\over{\partial t^2}}-2C_1 s^2 \epsilon {{\partial \tilde{W}}\over{\partial \xi}}
+ 4 cs C_2 \epsilon {{\partial U}\over{\partial \xi}} + 4 C_2 c^2
\epsilon^2 {{\partial^2 W}\over{\partial \xi^2}} }\cr \quad \cr
\displaystyle{ + sc(2C_2 - C_1) \epsilon^2 {{\partial^2 {\tilde
U}}\over{\partial \xi^2}} + 2(s^2 C_1 + 2 c^2 C_2) W +
2sc(2C_2-C_1) {\tilde U}  = 0,}\cr}
\label{fm2}
\end{equation}
\begin{equation}
\matrix{ \displaystyle{ {{\partial^2 {\tilde U}}\over{\partial t^2}}+2(C_1 c^2 + 4 C_2 s^2) \epsilon {{\partial U}\over{\partial \xi}}
+ 4 s c C_2 \epsilon {{\partial {\tilde W}}\over{\partial \xi}} -
4 C_2 s^2 \epsilon^2 {{\partial^2 {\tilde U}}\over{\partial
\xi^2}} }\cr \quad \cr \displaystyle{ + sc(2C_2 - C_1) \epsilon^2
{{\partial^2 W}\over{\partial \xi^2}} + 2(c^2 C_1 + 2 s^2 C_2)
{\tilde U} + 2sc(2C_2-C_1)  W}  = 0, \cr}
\label{fm3}
\end{equation}
\begin{equation}
\matrix{ \displaystyle{ {{\partial^2 {\tilde W}}\over{\partial t^2}}+2C_1 s^2 \epsilon {{\partial W}\over{\partial \xi}}
- 4 cs C_2 \epsilon {{\partial {\tilde U}}\over{\partial \xi}} + 4
C_2 c^2 \epsilon^2 {{\partial^2 {\tilde W}}\over{\partial \xi^2}}
}\cr \quad \cr \displaystyle{ + sc(2C_2 - C_1) \epsilon^2
{{\partial^2 U}\over{\partial \xi^2}} + 2(s^2 C_1 + 2 c^2 C_2)
{\tilde W} + 2sc(2C_2-C_1) U  = 0. }\cr}
\label{fm4}
\end{equation}
\noindent
Then we introduce operator approximation for $\ddot u_n$ and $\ddot u_{n+1}$ in corresponding equations using previous Taylor expansions and
expansion of dispersion relation according to
\begin{equation}
\left \{
\matrix{\displaystyle{ \ddot u_n \simeq -\omega^2 U = -\Omega^2 U + \nu \epsilon {{\partial {\tilde U}}\over{\partial \xi}} + \mu \epsilon^2
  {{\partial^2 U}\over{\partial \xi^2}} \dots
 }\cr
\displaystyle{\ddot u_{n+1} \simeq -\omega^2 {\tilde U} = -\Omega^2 {\tilde U} - \nu \epsilon {{\partial  U}\over{\partial \xi}} + \mu \epsilon^2
  {{\partial^2 {\tilde U}}\over{\partial \xi^2}} + \dots
}\cr} \right.
\label{opun}
\end{equation}
Substitution of (\ref{fm3}), (\ref{fm4}) into starting nonlinear
equation (\ref{f9}) leads to four nonlinear partial  differential
equations (NPDE) with respect to functions $W$, $\tilde{W}$, $U$, $\tilde{U}$.
To reduce the order of this system one can express the functions
$U$, $\tilde{U}$ via $W$, $\tilde{W}$ using the equations
for $U$ and $\tilde{U}$, and taking into account the terms up to
third degree. As this takes place, we have to replace the nonlinear terms
with respect $U$, $\tilde{U}$ by their expressions via $W$, $\tilde{W}$
obtained from (\ref{fm1}), (\ref{fm3}), (\ref{fm4}).
Then NPDE for $U$ and $\tilde{U}$ provide two independent linear relations
relative these functions that can be solved to
 obtain $U$ and ${\tilde U}$ versus $W$ and ${\tilde W}$.
Indeed truncation of the full
%Nonlinear Partial Differential Equations
NPDE at order 3 are used.
 At this step, one can notice that for $k = \pi/2$ or $k=k_0$ displacements
  $u_n$ and $w_n$ are of same order: we have longitudinal and transversal motions.
If measuring the length in units $l_z$ we have
$k \sim \pi/2:~ u_n \sim w_n$.
In strongly coupled general case, the relations can be written as:
\begin{equation}
U = A_1 W + A_2 {\tilde W} + {\cal F}_{12}(W,
{\tilde W}) +  {\cal F}_{13}(W, {\tilde W}) + A_3 \epsilon
{{\partial W}\over{\partial \xi}}  + A_4 \epsilon {{\partial
{\tilde W}}\over{\partial \xi}} + A_5 \epsilon^2 {{\partial^2
W}\over{\partial \xi^2}}+ A_6 \epsilon^2 {{\partial^2 {\tilde
W}}\over{\partial \xi^2}} + \dots,
\label{relU}
\end{equation}
\begin{equation}
{\tilde U} = B_1 W + B_2 {\tilde W} + {\cal F}_{22}(W,
{\tilde W}) +  {\cal F}_{23}(W, {\tilde W}) + B_3 \epsilon
{{\partial W}\over{\partial \xi}}  + B_4 \epsilon {{\partial
{\tilde W}}\over{\partial \xi}} + B_5 \epsilon^2 {{\partial^2
W}\over{\partial \xi^2}}+ B_6 \epsilon^2 {{\partial^2 {\tilde
W}}\over{\partial \xi^2}}  + \dots,
\label{relUt}
\end{equation}
with ${\cal F}_{ij}, i =1,2$ of degree $j$ in $W, {\tilde W}$.
$A_k, B_k$, $k=1, \dots, 6$ are given constants.
For example, if $k=\pi/2$, one has:
\begin{equation}
\left\{ \matrix{\displaystyle{
A_1= 0, A_2 =-{c \over s}, {\cal F}_{12}(W, {\tilde W})= D (W^2 + {\tilde W}^2),
}\cr\quad\cr
\displaystyle{
B_2= 0, B_1 =-{c \over s}, {\cal F}_{22}(W, {\tilde W})= -D (W^2 + {\tilde W}^2),
}\cr
\quad \cr \displaystyle{D = {{3 c q \rho_0 C_1}\over{2 s^4(C_1-2C_2)}}} \cr
}\right.
\label{relUUtpis2}
\end{equation}
\noindent
For $k=0$ or $k=\pi$, degeneracy of order occurs: $u$ or $w$ displacements dominate.
We have:
$$
\matrix{ k \sim 0:~ u_n \sim k w_n \sim \varepsilon w_n, \hbox{ we
suppose that } k \sim \varepsilon << 1 \cr
 k \sim \pi:~ w_n \sim (\pi-k) u_n
\sim \varepsilon u_n. \cr}
$$
Then modulating functions introduced in equations for $u_n$ and $w_n$
provide only 2 NPDE. In these two particular cases, one needs only
to consider equation for $u_n$ (or respectively $w_n$).
The relations, connecting $u_n$ and $w_n$ are obtained in the form
\begin{equation}
u_n =U =\varepsilon \left[\frac{c}{s} \frac{\partial
W}{\partial\xi} + \frac{c}{\rho_0 s^2}(2 - 3
c^2+3s^2\rho_0q-8\delta s^2) W^2 \right] + \dots
\label{rel0}
\end{equation}
for $k=0$ and
\begin{equation}
w_n= W =-\frac{sc}{2} \frac{1 -4\delta}{4\delta s^2 + c^2}
\varepsilon\frac{\partial U}{\partial \xi}+ \dots \label{relpi}
\end{equation}
for $k=\pi$.

%\begin{equation}
%\label{}
%\end{equation}

\section{Nonlinear partial equations}
Let us introduce new non-dimensional time by relation $\tau= \Omega t$,  where
$\Omega$ is linear frequency for considered normal mode.
After substitution of the expressions of the displacements via
modulating functions into
%planar Hamiltonian (\ref{f3}),
equations for $w_n$ and $w_{n+1}$,
one
can obtain corresponding nonlinear partial differential equations for
modulating functions
in the form:

\begin{equation} \matrix{ \displaystyle{ {{\partial^2 W}\over{\partial \tau^2}}+ {{f'(k)} \over \Omega^2} \epsilon {{\partial \tilde{W}}\over{\partial \xi}}
- {{f''(k)} \over \Omega^2}\epsilon^2 {{\partial^2 W}\over{\partial
\xi^2}}  +  W }\cr \quad \cr \displaystyle{ + {1\over{\Omega^2}}[
D_1 W^2 + D_2 W {\tilde W} + D_3 {\tilde W}^2 + D_4 W^3 + D_5 W^2
{\tilde W} + D_6 W {\tilde W}^2 + D_7 {\tilde W}^3] + \dots =
0,}\cr} \label{NPDEW}
\end{equation}

\begin{equation}\matrix{ \displaystyle{ {{\partial^2 {\tilde W}}\over{\partial \tau^2}}- {{f'(k)} \over \Omega^2} \epsilon {{\partial W}\over{\partial \xi}}
- {{f''(k)} \over \Omega^2} \epsilon^2 {{\partial^2 {\tilde
W}}\over{\partial \xi^2}} +   {\tilde W} }\cr \quad \cr
\displaystyle{+ {1\over{\Omega^2}}[- D_3 W^2 - D_2 W {\tilde W} -
D_1 {\tilde W}^2 + D_7 W^3 + D_6 W^2 {\tilde W} + D_5 W {\tilde
W}^2 + D_4 {\tilde W}^3 ] + \dots= 0, }\cr} \label{NPDEWt}
\end{equation}
where $D_j$,$j=1,\dots,7$ are constants depending on $C_1, C_2,
\rho_0, q, \theta_0$. For $k=\pi/2$, the expressions are given in
table~\ref{tableD}. They are not given for the general case
because of size of expressions. Numerical values for $K=K_0$, in re-scaled
nondimensional form corresponding to $\varepsilon =0.05$ are given in
table~\ref{tableD}.

\begin{table}[tb]
\caption{Values of constants $D_j$, $j=1,\dots,7$ for $k=\pi/2$
         and numerical values for $k=k_0$, corresponding to $\varepsilon= 0.05$.}
\begin{center}
\begin{tabular}{ccc}
\hline\hline
$j$ & $k=\pi/2$ & $k=k_0$ \\
\hline\hline
$1$&  $\displaystyle{{{(C_1-2C_2)D cs}/{C_1}}}$& 2.25\\
\hline
$2$&   0 &  -5.74\\
\hline
$3$& $\displaystyle{{{(C_1-2C_2)D cs}/{C_1}}}$ & 2.57\\
\hline
$4$ & -$\displaystyle{{{D(3C_1q\rho_0+C_1-2C_2)}/{C_1}}}$ & 1.89 \\
\hline
$5$ &  $\displaystyle{{{D(3C_1q\rho_0+C_1-2C_2)}/{C_1}}}$ & -1.14 \\
\hline
$6$ & -$\displaystyle{{{D(3C_1q\rho_0+C_1-2C_2)}/{C_1}}}$ & 2.42 \\
\hline
$7$  & ~~~~$\displaystyle{{{D(3C_1q\rho_0+C_1-2C_2)}/{C_1}}}$~~~~  & -1.14 \\
\hline
 \end{tabular}
\end{center}
\label{tableD}
\end{table}

%\begin{table}[tbh]
%\caption{Numerical values of constants $D_j$, $j=1,\dots,7$ for
%$k=k_0$, corresponding to $\varepsilon = 0.05$.}
%\begin{center}
%\begin{tabular}{ccc}
%\hline\hline
%$j$ & $D_j$ \\
%\hline\hline
%$1$&  $0.09$& \\
%\hline
%$2$&   $-0.23$ &  \\
%\hline
%$3$& $0.10$ & \\
%\hline
%$4$ & $0.08$ &  \\
%\hline
%$5$ &  $-0.05$ &  \\
%\hline
%$6$ & $0.10$ &  \\
%\hline
%$7$  & $-0.05$  &  \\
%\hline
% \end{tabular}
%\end{center}
%\label{tableD}
%\end{table}
% \begin{table}[tbh]
% \caption{Numerical values of constants $D_j$, $j=1,\dots,7$ for
% $k=k_0$, corresponding to $\varepsilon = 0.05$.}
% \begin{center}
% \begin{tabular}{ccc}
% \hline\hline
% $j$ & $D_j$ \\
% \hline\hline
% $1$&  $2.25$& \\
% \hline
% $2$&   $-5.74$ &  \\
% \hline
% $3$& $2.57$ & \\
% \hline
% $4$ & $1.89$ &  \\
% \hline
% $5$ &  $-1.14$ &  \\
% \hline
% $6$ & $2.42$ &  \\
% \hline
% $7$  & $-1.14$  &  \\
% \hline
%  \end{tabular}
% \end{center}
% \label{tableD}
% \end{table}

Again, degeneracy of cases $k=0$ or $k=\pi$ leads to only one equation obtained from equation for $w_n$ as:
\begin{equation}
\left\{ \matrix{ \displaystyle{ {{\partial^2
\tilde{W}}\over{\partial \tau^2}} - \lambda_1
\varepsilon^2 {{\partial^2 \tilde{W}}\over{\partial \xi^2}} +
\tilde{W} }
\displaystyle{
+  \alpha_2 {\tilde{W}}^3= 0,}
 \cr \quad
\cr \displaystyle{ \lambda_1 = {{C_1 \cos(\theta_0) + 4 c^2 C_2}
\over {4s^2 C_1}},} \cr \quad
 \cr
 \displaystyle{
\alpha_2 = {1 \over{\rho_0^2}} [2(5 c^2 -4)c^2/s^2 + {{14}\over 3} \rho_0^2 q^2 s^2 -12 \rho_0
q c^2  + 32 \delta c^2] ,
} \cr} \right.
\label{PDE0}
\end{equation}
for $k=0$ and
\begin{equation}
 \left\{ \matrix{ \displaystyle{ {{\partial^2
\tilde{U}}\over{\partial \tau^2}} - \lambda_2 \epsilon^2 {{\partial^2
\tilde{U}}\over{\partial \xi^2}} + \tilde{U} +  \alpha_1
{\tilde{U}}^2 +
 \alpha_2{\tilde{U}}^3 = 0,}\cr \quad \cr
\displaystyle{ \lambda_2 = -{{C_1 \cos(\theta_0) c^2 + 20 c^2 s^2 C_2
%+ (32-48c^2)s^2 \delta^2
}\over{4(c^2 + 4 s^2 \delta)^2}},}\cr \quad \cr \displaystyle{
\alpha_1 = {3 \over c}{{c^2(s^2-c^2 q\rho_0) + 4
s^2(1-4c^2)\delta}\over{c^2 + 4 s^2 \delta}},} \cr \quad \cr
\displaystyle{ \alpha_2 = - { {2 [ 3 c^2 s^2(5 c^2-1) + 18 c^4 s^2
q \rho_0 - 7 c^6 q^2 \rho_0^2 -12 \delta s^2(48 c^4 - 24
c^2+1)]}\over {3 c^2 (c^2 + 4 s^2 \delta)}}, }\cr}\right.
\label{PDEpi}
\end{equation}
for $k=\pi$.
In the last NPDE, nonlinear terms involving spatial derivatives can be neglected
because they are of higher orders by $\epsilon$.

All these NPDE have been given in dimensional form. They could be re-scaled
according to physical orders of the
 different variables (e.g. replacing the variables $U, {\tilde U}, W, {\tilde W}$ by $\varepsilon\rho_0 U$, $\varepsilon\rho_0 {\tilde U}$,
$\varepsilon\rho_0 W$, $\varepsilon\rho_0
\tilde{W}$). This has been done for numerical applications in the cases
$k\approx0$, $k\approx\pi$ and $k= k_0$ below. Such re-scaling makes the
orders of nonlinear terms to be similar to those of the terms containing
second spatial derivatives.

\section{Transition to complex variables}
The NPDE
%Nonlinear Partial Differential Equations
 of previous section
%equation (\ref{eq25})
can be written under first
order by time if setting

\begin{equation}
\matrix{ k\sim 0,~~ \Psi(\xi, \tau) = (W'
%\dot{W}
+ i {W}),~~ \Psi^\star(\xi, \tau) =
(
%\dot{W}
W'
 - i {W}), \cr\quad \cr
k\sim\pi,~~ \Psi(\xi, \tau) = (
%\dot{U}
U'
+ i {U}),~~
\Psi^\star(\xi, \tau) = (
%\dot{U}
U'
 - iU), \cr \quad \cr
k\sim\pi/2 \hbox{ or } k_0,~~ \left \{ \matrix{
 \Psi_1(\xi, \tau) = (
 %\dot{W}
W'
 + i W),~~
 \Psi_1^\star(\xi, \tau) =
 (
 %\dot{W}
 W'
  - i  W), \cr \quad \cr
 \Psi_2(\xi, \tau) = (
 %\dot{\tilde{W}}
 {\tilde{W}}'
 + i\tilde{W}),~~
 \Psi_2^\star(\xi, \tau) = (
 %\dot{\tilde{W}}
{\tilde{W}}'
 -i\tilde{W}).
  \cr}\right. \cr} \label{complex}
\end{equation}
We obtain two NPDE, for $k=0$:
\begin{equation}\left\{
\matrix{ \displaystyle{-i {{\partial \Psi}\over{\partial \tau}} =
\Psi  - {\lambda_1 \over 2} \epsilon^2 {{\partial^2
}\over{\partial \xi^2}}(\Psi-\Psi^\star)-{\alpha_2 \over 8}
(\Psi-\Psi^\star)^3+\dots }\cr \quad \cr \hbox{ Conjugate equation
},\cr}\right. \label{eq28}
\end{equation}
and  $k=\pi$ :
\begin{equation}\left\{
\matrix{ \displaystyle{-i {{\partial \Psi}\over{\partial \tau}} =
\Psi  - {\lambda_2 \over 2} \epsilon^2 {{\partial^2
}\over{\partial \xi^2}}(\Psi-\Psi^\star)-{i\alpha_1 \over 4}
(\Psi-\Psi^\star)^2-{\alpha_2 \over 8} (\Psi-\Psi^\star)^3+\dots
}\cr \quad \cr \hbox{ Conjugate equation },\cr}\right.
\label{eq28b}
\end{equation}
In general case one can obtain four NDPE, e.g. for $k=\pi/2$ they have
linear part
\begin{equation}
\left \{ \matrix{ \displaystyle{-i {{\partial
\Psi_1}\over{\partial \tau}} = \Psi_1 + {{f'(k)}\over{\Omega^2}}
\epsilon {{\partial^2}\over{\partial \xi^2}}(\Psi_2-\Psi_2^\star)
- {{f"(k)}\over{\Omega^2}} \epsilon^2 {{\partial^2}\over{\partial
\xi^2}}(\Psi_1-\Psi_1^\star),  } \cr \quad \cr \hbox{ Conjugate
equation,} \cr \quad \cr \displaystyle{-i {{\partial
\Psi_2}\over{\partial \tau}} = \Psi_2 - {{f'(k)}\over{\Omega^2}}
\epsilon {{\partial^2}\over{\partial \xi^2}}(\Psi_1-\Psi_1^\star)
- {{f"(k)}\over{\Omega^2}} \epsilon^2 {{\partial^2}\over{\partial
\xi^2}}(\Psi_2-\Psi_2^\star),  }\cr \quad \cr \hbox{Conjugate
equation}.\cr}\right. \label{eq29}
\end{equation}
Let us introduce the changes of variables
\begin{equation}
\tau_0=\tau,~~ \tau_1 = \varepsilon \tau_0,~~    \tau_2 = \varepsilon^2 \tau_0, \dots
\label{times}
\end{equation}
Setting
$\psi = \phi e^{i \tau}$ (for $k=0$, $\pi$) and $\psi_j = \phi_j e^{i
\tau}$, $j=1,2$ (for $k=\pi/2$) we use farther the power expansions
\begin{equation}
\phi = \varepsilon (\phi_0 + \varepsilon \phi_1 + \varepsilon^2
\phi_2 + \dots) \label{expansion}
\end{equation}
and respectively
\begin{equation}
\phi_j = \varepsilon (\phi_{j0} + \varepsilon\phi_{j1} +
\varepsilon^2\phi_{j2} + \dots )\label{expansion1}
\end{equation}
Then we obtain at order $\varepsilon^0$ the relations:
\begin{equation}
k=0:~  -i \displaystyle{{{\partial \phi_0}\over{\partial \tau_0}} = 0, }
\label{eq30}
\end{equation}
\begin{equation}
k=\pi:~ -i \displaystyle{{{\partial \phi_0}\over{\partial \tau_0}} = 0, }
\label{eq31}
\end{equation}
\begin{equation}
k = k_0:~ -i \displaystyle{{{\partial \phi_{10}}\over{\partial \tau_0}} = 0},~~
-i \displaystyle{{{\partial \phi_{20}}\over{\partial \tau_0}} = 0, }
\label{eq32}
\end{equation}
It means that for $k=0$, $\pi$, $\phi_0 = \phi_0(\xi,\tau_1,\tau_2)$ and for
$k=k_0$, $\phi_{1,0} = \phi_{1,0}(\xi,\tau_1,\tau_2)$, $\phi_{2,0} =
\phi_{2,0}(\xi,\tau_1,\tau_2)$.
For $k=\pi/2$ presence of first space derivatives lead to different problem.

At order $\varepsilon^1$ we have the equations
\begin{equation}
k = 0, \pi:~ \displaystyle{{{\partial \phi_{1}}\over{\partial \tau_0}} +
{{\partial \phi_{0}}\over{\partial \tau_1}} =0,  }
\end{equation}
\begin{equation}
k =  k_0:~ \displaystyle{{{\partial \phi_{1,1}}\over{\partial \tau_0}} +
 {{\partial \phi_{1,0}}\over{\partial \tau_1}}
 % -
 %\left(
 %{{\partial
 %\phi_{1,0}}\over{\partial
 %\xi}}
 %-e^{-2i\tau_0}\frac{\partial\phi^\star_{1,0}}{\partial\xi}
 %\right)
  = 0,  }
\end{equation}
\begin{equation}
 \frac{\partial\phi_{2,1}}{\partial\tau_0}+
 \frac{\partial\phi_{2,0}}{\partial\tau_1}
%-\left(
 %\frac{\partial\phi^\star_{2,0}}{\partial\xi}
 %-e^{-2i\tau_0}\frac{\partial\phi^\star_{2,0}}{\partial\xi}\right)
 =0
\end{equation}

Conditions of absence of resonances lead to equations
\begin{equation}
k = 0,~\pi:~~ \displaystyle{{{\partial \phi_{0}}\over{\partial \tau_1}} =0,  }
\end{equation}
\begin{equation}
k =  k_0:~~\displaystyle{ {{\partial \phi_{1,0}}\over{\partial
\tau_1}}
%-{{\partial \phi_{1,0}}\over{\partial \xi}}
 =0,
 %}
%\end{equation}
 %\begin{equation}
 \frac{\phi_{2,0}}{\partial\tau_1}
 %-\frac{\partial\phi_{2,0}}{\partial\xi}
 =0. }\hfill \qquad
 \end{equation}

 Therefore $\phi_0 = \phi_0(\xi, \tau_2,\tau_3, \dots)$ for $k=0, \pi$ as for
the case $k=k_0$, we can write that $\phi_{j,0}= \phi_{j,0}(\xi, \tau_2,
\dots)$, $j=1,2$.

At order $\varepsilon^2$ the conditions of absence of resonance terms lead to
final complex equations in main approach:
\begin{equation}
\matrix{ \displaystyle{ k=0,\pi:~~ -i {{\partial
\phi_0}\over{\partial \tau_2}} + \beta{{\partial^2
\phi_{0}}\over{\partial \xi^2}} + \alpha \mid \phi_0\mid^2
\phi_0 =0,}\cr \displaystyle{k=0: \alpha = -{{3 \alpha_2}\over{8}} , \beta=
\displaystyle{{\lambda_1/2}},}\cr \displaystyle{k=\pi, \alpha = -{{3 \alpha_2}\over{8}}+
{{5 \alpha_1^2}\over{12}}, \beta = \displaystyle{{\lambda_2/2}}} .\cr}
 \label{eq33 }
\end{equation}
which is Nonlinear Schr{\"o}dinger Equation (NSE)
for $k=0$ and $k=\pi$ respectively. For $k=k_0$, the resonant equations have the general form:
\begin{equation}
\matrix{
 \displaystyle{   -i {{\partial \phi_{1,0}}\over{\partial \tau_2}}
-{\lambda_2 \over 2} {{\partial^2 \phi_{10}}\over{\partial \xi^2}}
+ P_1 {\mid \phi_{10} \mid}^2\phi_{10} + P_2 {\mid
\phi_{10} \mid}^2\phi_{20} + P_5 {\mid
\phi_{20} \mid}^2\phi_{10}
}
\cr
+ P_6 {\mid
\phi_{20} \mid}^2\phi_{20}+ P_4 \phi_{10}^2 \overline{\phi_{20}} + P_5 \phi_{20}^2 \overline{\phi_{10}} =0,\cr
} \label{eq34}
\end{equation}
\begin{equation}
\matrix{
 \displaystyle{   -i {{\partial \phi_{2,0}}\over{\partial \tau_2}}
%+ \beta_{21} {{\partial^2 \phi_{10}}\over{\partial \xi^2}}
 -{\lambda_2 \over 2}
{{\partial^2 \phi_{20}}\over{\partial \xi^2}} + P_1 {\mid \phi_{20} \mid}^2\phi_{20} + P_2 {\mid
\phi_{20} \mid}^2\phi_{10} + P_5 {\mid
\phi_{10} \mid}^2\phi_{20} }\cr+ P_6 {\mid
\phi_{10} \mid}^2\phi_{10}+ P_4 \phi_{20}^2 \overline{\phi_{10}} + P_5 \phi_{10}^2 \overline{\phi_{20}} =0
=0. \cr} \label{eq35}
\end{equation}
with
 $P_i, i=1,6$ constants derived from NPDE (\ref{NPDEW}) and (\ref{NPDEWt}):
\begin{eqnarray}
P_1 &=& {3 \over 8}D_4 -{5 \over{12}} D_1^2 + {5 \over{24}} D_2 D_3, \nonumber \\
P_2 &=& {1 \over 8}D_5 +{1 \over{2}} D_3^2 - {5 \over{12}} D_1 D_2  + {1 \over{12}} D_2^2, \nonumber \\
P_3 &=& -{1 \over{12}} D_3^2 - {5 \over{24}} D_1 D_2 + {1 \over{8}} D_2^2 + {1 \over{8}} D_5, \nonumber \\
P_4 &=& -{1 \over{24}} D_1 D_2 - {1 \over{8}} D_2^2 + {1 \over{12}} D_1 D_3 + {1 \over{4}} D_2 D_3 + {1 \over{8}} D_6 , \nonumber \\
P_5 &=& {3 \over{8}} D_7 - {5 \over{24}} D_2 D_3 + {5 \over{12}} D_1 D_3,\nonumber
\\
P_6 &=& -{1 \over{2}} D_1 D_3 - {1 \over{12}} D_2^2 + {1 \over{4}} D_6 + {1 \over{6}} D_2 D_3 + {1 \over{4}} D_1 D_2. \nonumber
\end{eqnarray}

\section{Soliton-like solutions}
As it is known, the type of soliton-like (breather) solutions
depends strongly on relationship between the signs of constants
$\alpha$ and $\beta$ of NSE. When these
signs are similar, NSE admits the envelope solitons. In opposite
case, NSE admits "dark" solitons. From this point of view there is
a significant difference between two systems of zigzag parameters
introduced above.

For the case $k \sim 0$, and both values of
$\delta$, the signs of  $\alpha$ and $\beta$ are the same ($<0$) (see
numerical values in table~\ref{tableab}).

\begin{table}[tb]
\caption{Values of $\alpha, \beta$ versus $\delta$ for $k \sim 0$ or $\pi$.}
\begin{center}
\begin{tabular}{ccccc}
\hline\hline
$k$ & $\delta$ & $\alpha$ & $\beta$ \\
\hline\hline
$0$& $0.019$ & $0.053$ & $-0.07$\\
\hline
$0$& $0.078$ & $0.054$ & $-0.05$\\
\hline
$\pi$&  $0.019$ & $0.035$ & $0.021$ \\
\hline
$\pi$& $0.078$ & $0.033$ & $-0.23$\\
\hline
 \end{tabular}
\end{center}
\label{tableab}
\end{table}

%\begin{table}[tbh]
%\caption{Values of $\alpha, \beta$ versus $\delta$ for $k \sim 0$
%or $\pi$.}
%\begin{center}
%\begin{tabular}{ccccc}
%\hline\hline
%$k$ & $\delta$ & $\alpha$ & $\beta$ \\
%\hline\hline
%$0$& $0.019$ & $8.92$ & $-0.07$\\
%\hline
%$0$& $0.078$ & $9.13$ & $-0.05$\\
%\hline
%$\pi$&  $0.019$ & $5.86$ & $0.021$ \\
%\hline
%$\pi$& $0.078$ & $5.64$ & $-0.23$\\
%\hline
% \end{tabular}
%\end{center}
%\label{tableab}
%\end{table}

For the case ($k \sim \pi$, $\delta=0.019$) the dispersion
curve has a form similar to Figure \ref{fg02} (a). In such a case
the signs of $\alpha$ and $\beta$ are the same ($>0$). Then one
can obtain the the envelope solitons (or breathers) as particular
solutions:
\begin{eqnarray}
\phi_0(\xi,\tau_2)&=&(2A/\alpha_s)^{1/2}\exp(iv\xi/2\sqrt{\beta_s}+i\omega\tau_2)
\nonumber \\
&&\times\mbox{sech}[A^{1/2}(\xi/\sqrt{\beta_s}+v\tau_2)], \label{eq36}
\end{eqnarray}
where $\omega=v^2/4-A$.

For the second system ($k \sim \pi$, $\delta=0.078$) the
dispersion curve has a form similar to Figure \ref{fg02} (b)
(optic branch) and  dark solitons exist.
% old figure 2!
% old figure 2!  \begin{figure}[tb]
% old figure 2!  \begin{center}
% old figure 2!  \includegraphics[angle=0, width=1\linewidth]{fg02.eps}
% old figure 2!  \end{center}
% old figure 2!  \caption{\label{fg02}\protect
% old figure 2!          Shape of dispersion curves for $k$ close to $\pi$.
% old figure 2!          }
% old figure 2!  \end{figure}

In the case $k \sim k_0$,  breather or dark soliton can exist near
minimal frequency. This fact has been confirmed by computer
simulation. Looking for particular solutions of the form
$\phi_{2,0} = \gamma \phi_{10}$, we obtain only two possible
values for $\gamma$: $\gamma = \pm 1$. The two coupled complex
NPDE (\ref{eq34}) and (\ref{eq35}) can be reduced to only one
Schr\"odinger equation of the form
\begin{equation}
 \displaystyle{   -i {{\partial \phi_{1,0}}\over{\partial \tau_2}}
+ \beta {{\partial^2 \phi_{10}}\over{\partial \xi^2}}
+ \alpha {\mid \phi_{10} \mid}^2\phi_{10}
 =0,
}
\label{schrk0}
\end{equation}
with $\beta = -\lambda_2/2$, $\alpha = P_1 + P_4 + P_6 + \gamma
(P_2 + P_3 + P_5)$. Numerical values are given in
table~\ref{tableabk0}. The breathers exist for $\gamma = 1$ and
dark solitons for $\gamma=-1$.

%\begin{table}[tbh] \caption{Values of $\alpha, \beta$ versus
%$\delta$ for $k \sim k_0$, $\gamma = \pm 1$.}
%\begin{center}
%\begin{tabular}{ccccc}
%\hline\hline
%$\gamma$ & $\delta$ & $\alpha$ & $\beta$ \\
%\hline\hline
%$+1$& $0.078$ & $-0.004$ & $0.28$\\
%\hline
%$-1$& $0.078$ & $0.099$ & $0.28$\\
%\hline
%\hline
% \end{tabular}
%\end{center}
%\label{tableabk0}
%\end{table}

\begin{table}[tb]
\caption{Values of $\alpha, \beta$ versus $\delta$ for $k \sim
k_0$, $\gamma = \pm 1$.}
\begin{center}
\begin{tabular}{ccccc}
\hline\hline
$\gamma$ & $\delta$ & $\alpha$ & $\beta$ \\
\hline\hline
$+1$& $0.078$ & $-0.010$ & $0.28$\\
\hline
$-1$& $0.078$ & $2.47$ & $0.28$\\
\hline \hline
 \end{tabular}
\end{center}
\label{tableabk0}
\end{table}

\section{Numerical simulations}
To check the validity of assumption made in the analytical study,
we have undertaken a numerical treatment of the breathers
existence as well as their stability in free motions, under
collisions and thermal perturbations.

While numerical modeling the breathers and their dynamics,
we consider the following system of equations corresponding
to Hamiltonian (\ref{f3}):
\begin{equation}
m\ddot{u}_n=-\frac{\partial H}{\partial u_n},~~~
m\ddot{v}_n=-\frac{\partial H}{\partial v_n},~~~
m\ddot{w}_n=-\frac{\partial H}{\partial w_n} \label{n1}
\end{equation}
for $n=1,2,...,N$.

We use initial conditions in agreement with approximate analytical solution.
Because of small difference from exact solution, there will be a phonon
radiation. For its absorption, the viscous friction is introduced at the end
of the chain. As it was mentioned above, we deal with two systems of parameters
for PE crystal. One of them is\cite{p19}:
\begin{eqnarray}
&&m=14m_p,~~
p_1=8.37\mbox{kJ/mol},~~
p_2=1.675\mbox{kJ/mol},\nonumber \\
&&p_3=6.695\mbox{kJ/mol},~~
\Lambda=130.122\mbox{kJ/mol},
 \label{n2} \\
&&\theta_0=113^\circ,~~D=334.72\mbox{kJ/mol},\nonumber\\
&&q=1.91\mbox{\AA}^{-1},~~
\rho_0=1.53\mbox{\AA}, \nonumber
\end{eqnarray}
where $m_p$ is the proton mass.
Second system of parameters which is used in
\cite{p20} differs with more high value of parameter
\begin{equation}
\Lambda=529\mbox{kJ/mol}. \label{n3}
\end{equation}
When using the first system of parameters (\ref{n2}), a geometric nonlinearity plays
a crucial role in nonlinear dynamics of the PE chain. In the second
case (\ref{n3}) a physical nonlinearity becomes more essential. Moreover, the dispersion
curves for these two cases have different view. Respectively, the optic breathers
will have different view and different regions of existence in parametric space
(Fig.  \ref{fg02}).
Therefore we study numerically their properties for both systems of parameters.

\begin{figure}[tb]
\begin{center}
\includegraphics[angle=0, width=1\linewidth]{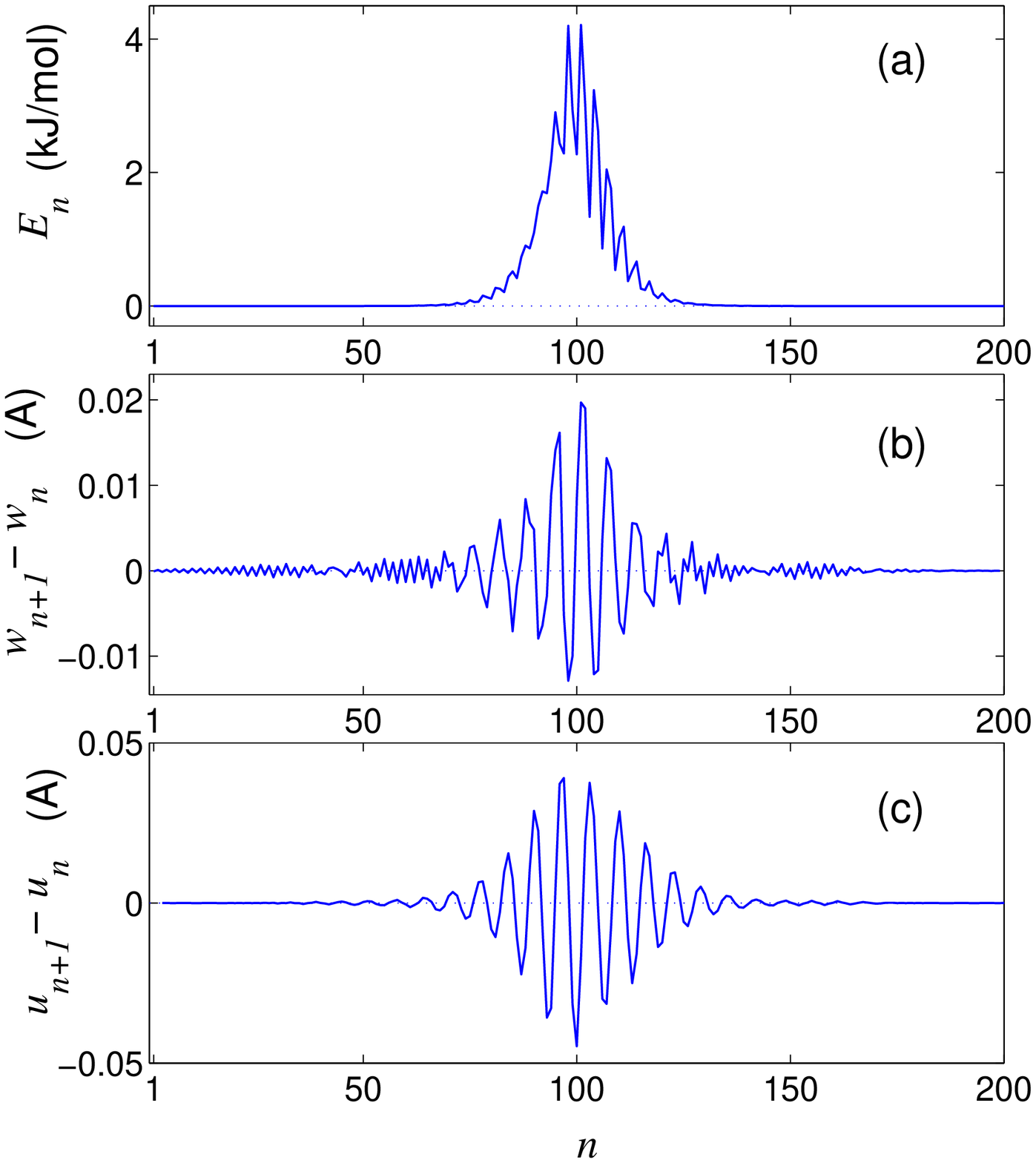}
\end{center}
\caption{\label{fig4a}\protect
        Optical breather in PE chain (parameter $\gamma=529$ kJ/mol).
        The distribution along the chain energy $E_n$ (a), longitudinal
        $w_{n+1}-w_n$ (b) and transversal $v_{n+1}-v_n$ (c) relative
        displacements of the chain segments
        is shown (breather energy  $E=56.06$ kJ/mol, frequency
        $\omega=926$ cm$^{-1}$, velocity $s=0.15$ (1146 m/s).
        }
\end{figure}

\begin{figure}[tb]
\begin{center}
\includegraphics[angle=0, width=1\linewidth]{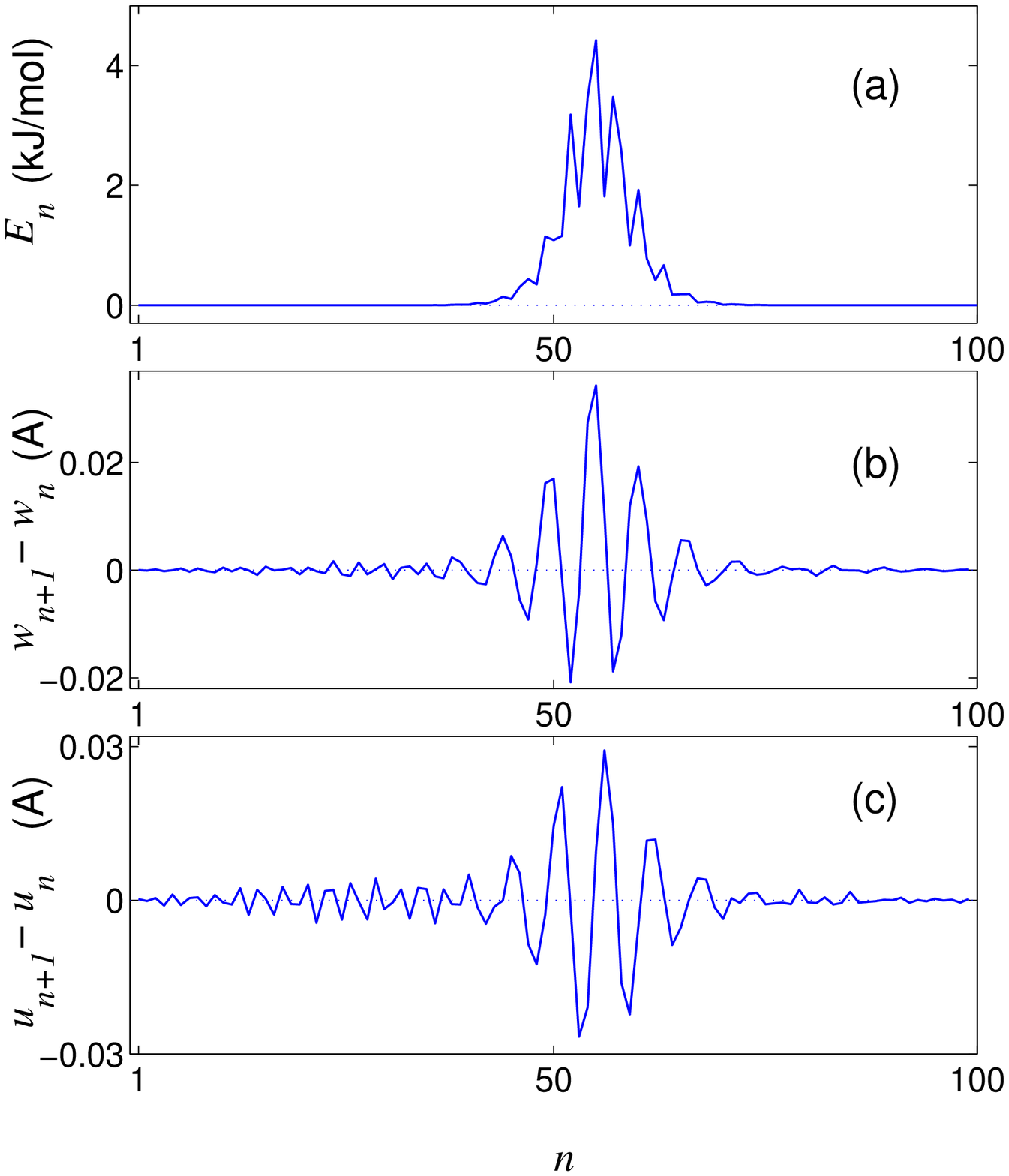}
\end{center}
\caption{\label{fig5a}\protect
        Optical breather in PE chain (parameter $\gamma=130.122$ kJ/mol).
        The distribution along the chain energy $E_n$ (a), longitudinal
        $w_{n+1}-w_n$ (b) and transversal $v_{n+1}-v_n$ (c) relative
        displacements of the chain segments
        is shown (breather energy  $E=31.1$ kJ/mol, frequency
        $\omega=909$ cm$^{-1}$, velocity $s=0.48$ (3888 m/s).
        }
\end{figure}

\begin{figure}[tb]
\begin{center}
\includegraphics[angle=0, width=1\linewidth]{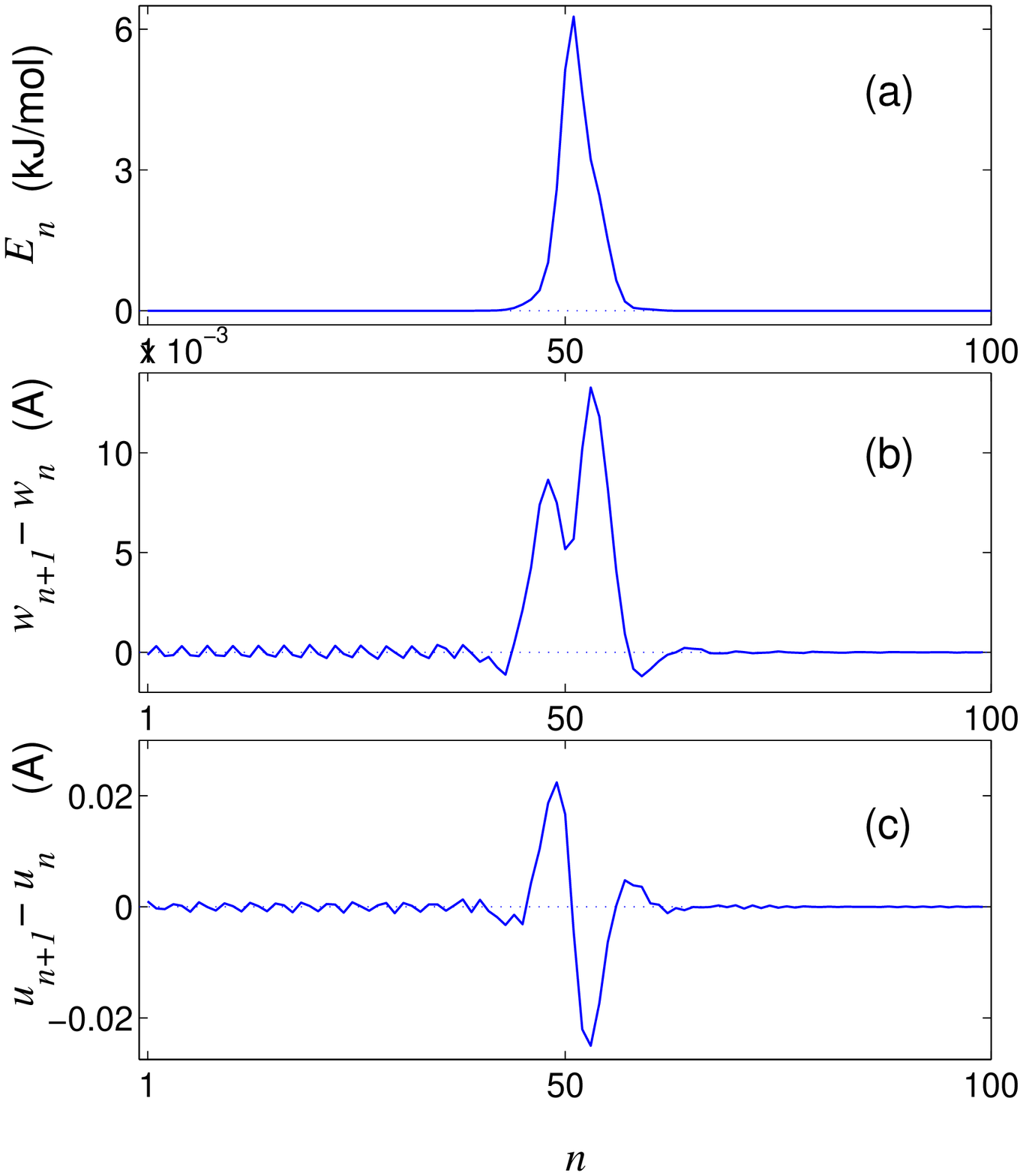}
\end{center}
\caption{\label{fig6a}\protect
        Optical breather in PE chain (parameter $\gamma=130.122$ kJ/mol).
        The distribution along the chain energy $E_n$ (a), longitudinal
        $w_{n+1}-w_n$ (b) and transversal $v_{n+1}-v_n$ (c) relative
        displacements of the chain segments
        is shown (breather energy  $E=28.77$ kJ/mol, frequency
        $\omega=830$ cm$^{-1}$, velocity $s=0.16$ (1296 m/s).
        }
\end{figure}

\begin{figure}[tb]
\begin{center}
\includegraphics[angle=0, width=.8\linewidth]{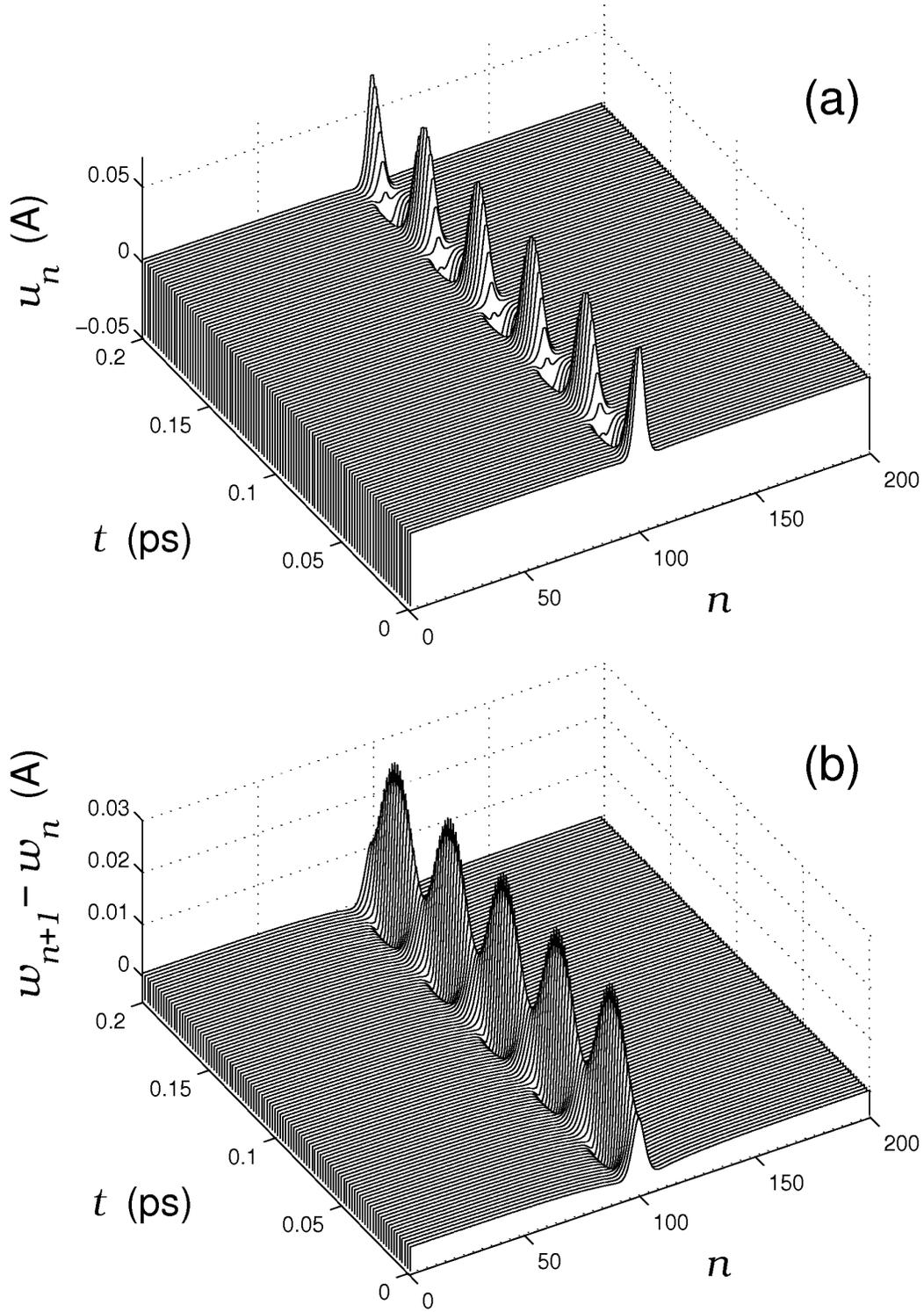}
\end{center}
\caption{\label{fg03}\protect
        Periodic change of transversal $u_n$ and relative longitudinal displacements
        of zigzag $w_{n+1}-w_n$ in the localization region of optic breathers under
        parameters (\ref{n2}), $N=200$. Frequency of the breather $\omega=820.5$
        cm$^{-1}$ is slightly lower than gap frequency.
        }
\end{figure}

\begin{figure}[tb]
\begin{center}
\includegraphics[angle=0, width=.8\linewidth]{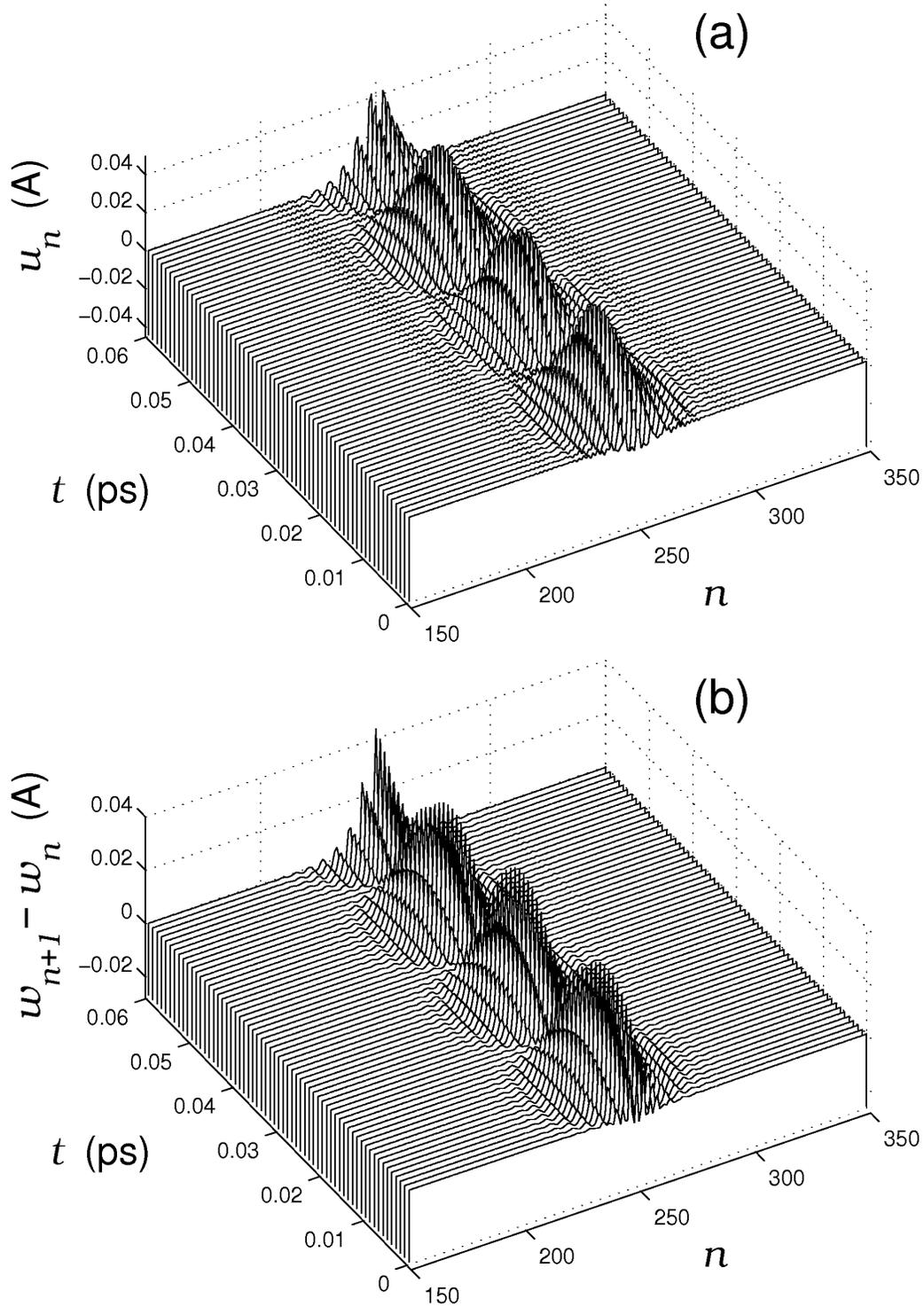}
\end{center}
\caption{\label{fg04}\protect
        Periodic change of transversal $u_n$ and relative longitudinal displacements
        of zigzag chain $w_{n+1}-w_n$ in the localization region of optic breathers
        under parameter (\ref{n3}), $N=500$. Frequency of breather $\omega=928.4$
        cm$^{-1}$ is slightly lower than frequency corresponding to low boundary
        of gap.
        }
\end{figure}

Numerical integration of the equation of motion (\ref{n1}) has confirmed
that in accordance with analytical study, the optic breathers exist for
both systems of parameters  (\ref{n2}) and (\ref{n3}) near low boundaries
of the frequencies of optic phonons.
Typical distribution of relative displacements in the localization region
of planar breather is presented in Fig. \ref{fig4a}, \ref{fig5a}, \ref{fig6a}.
One can see that shift of frequency (see Fig. \ref{fig4a} and \ref{fig5a})
leads to  narrowing of breather. Analogous effect is achieved also when
increasing the intensity of excitation -- see Fig. \ref{fig6a}.
Characteristics of breathers for the systems of parameters (\ref{n2}), (\ref{n3}).
are essentially
different -- see Fig. \ref{fg03} and \ref{fg04}. In the localization regions
of breathers the local change of valence angles is accompanied by local
extension of zigzag (average values of relative longitudinal displacements
$w_{n+1}-w_n$ are positive).
%---------------------------- Fig. 4a ------------------------------------
%---------------------------- Fig. 4a ------------------------------------
%---------------------------- Fig. 5a ------------------------------------
%---------------------------- Fig. 5a ------------------------------------
%---------------------------- Fig. 6a ------------------------------------
%---------------------------- Fig. 6a ------------------------------------

%---------------------------- Fig. 3 ------------------------------------
%---------------------------- Fig. 3 ------------------------------------
%---------------------------- Fig. 4------------------------------------
%---------------------------- Fig. 4 ------------------------------------

To consider an interaction of breathers with thermal vibrations of
the chain, $N_0$ segments of the chain were inserted from every
side into thermal bath with temperature T. Then the Langevin
equations of motion
\begin{eqnarray}
m\ddot{u}_n&=&-\frac{\partial H}{\partial u_n}+\xi_n-\Gamma_nm\dot{u}_n,\nonumber \\
m\ddot{v}_n&=&-\frac{\partial H}{\partial v_n}+\eta_n-\Gamma_nm\dot{u}_n,\label{n4}\\
m\ddot{w}_n&=&-\frac{\partial H}{\partial w_n}+\zeta_n-\Gamma_nm\dot{u}_n,\nonumber
\end{eqnarray}
where the Hamiltonian of the system H is given by Eq. (\ref{f3}),
$\xi_n$, $\eta_n$, and $\zeta_n$ are random normally distributed forces describing
the interaction of $n$th molecule with a thermal bath,
coefficient of friction $\Gamma_n=0$, forces $\xi_n$, $\eta_n$, $\zeta_n=0$ for
$N_0<n\le N-N_0$ and $\Gamma_n=\Gamma$ for $n\le N_0$ and $N-N_0<n\le N$.
Coefficient of friction $\Gamma=1/t_r$ , where $t_r$ is the relaxation
of the velocity of the molecule. The random forces $\xi_n$, $\eta_n$, and $\zeta_n$
have correlation functions
\begin{eqnarray*}
\langle\xi_n(t_1)\xi_l(t_2)\rangle=\langle\eta_n(t_1)\eta_l(t_2)\rangle=
\langle\zeta_n(t_1)\zeta_l(t_2)\rangle\nonumber\\
=2m\Gamma k_BT\delta_{nl}\delta(t_1-t_2),\\
\langle\xi_n(t_1)\eta_l(t_2)\rangle=\langle\xi_n(t_1)\zeta_l(t_2)\rangle=
\langle\eta_n(t_1)\zeta_l(t_2)\rangle=0, \\
1\le n,l \le N_0,~~~N-N_0< n,l\le N,
\end{eqnarray*}
where $k_B$ is Boltzmann's constant and $T$ is the temperature of heat bath.

The system (\ref{n4}) was integrated numerically by the standard
forth-order Runge-Kutta method with a constant step of integration
$\Delta t$. Numerically, the $\delta$-function was represented as
$\delta(t)=0$ for $|t|>\Delta t/2$ and $\delta(t)=1/\Delta t$ for $|t|\le\Delta t/2$,
i.e. the step of numerical integration corresponded to the correlation
time of the random force. In order to use the Langevin equation,
it is necessary that $\Delta t\ll t_r$. Therefore, we chose
$\Delta t=0.001$ ps and relaxation time $t_r=0.1$ ps.

\begin{figure}[tb]
\begin{center}
\includegraphics[angle=0, width=1\linewidth]{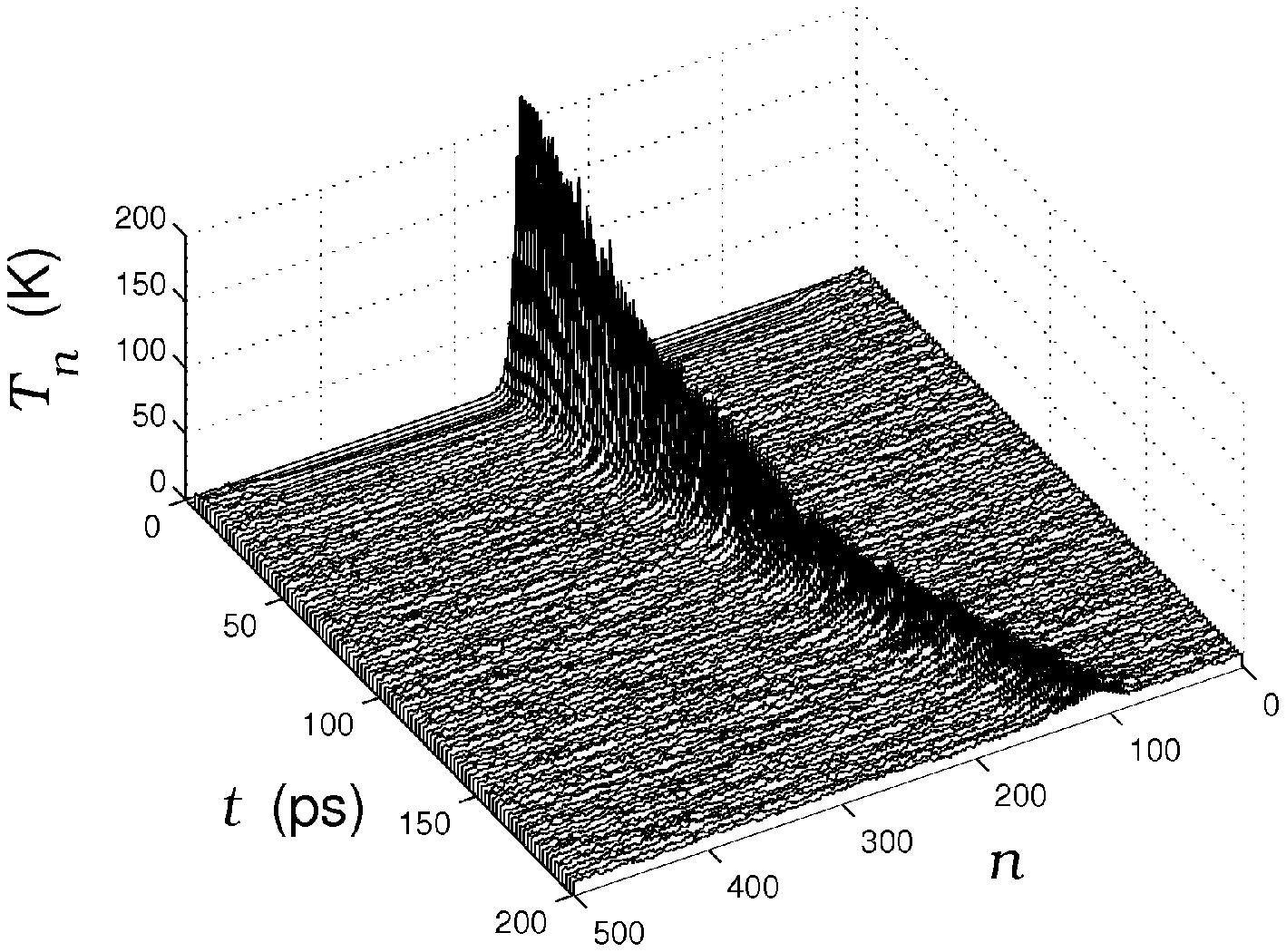}
\end{center}
\caption{\label{fg05}\protect
        Breaking of the breather (frequency $\omega=928.4$ cm$^{-1}$) in thermalized
        chain ($N=500$, $T=10$ K, $\Lambda=529$ kJ/mol).
        Temporal dependence of current local magnitudes of temperature
        (kinetic energy of chain segments) $T_n$ is presented.
        }
\end{figure}

To avoid an effect of friction coefficient on the behavior of breather,
it was isolated from heat bath. For this, the stationary breather was situated
at the center of the chain with $N_0=50$.
In such a case, the breather can interact only with thermal phonons arising
at the ends of the chain, which are connected with heat bath. The numerical
integration of the equations (\ref{n4}) has shown that, contrary to isolated
chain, the breathers in thermalized chain have a finite time of life. However,
this time is large enough to provide a significant role of the breathers in
different physical processes. Breaking of the breathers in thermalized chain is
shown at Fig.  \ref{fg05} and \ref{fg12} (a).
%---------------------------- Fig. 5 ------------------------------------
%---------------------------- Fig. 5 ------------------------------------

\begin{figure}[tb]
\begin{center}
\includegraphics[angle=0, width=.9\linewidth]{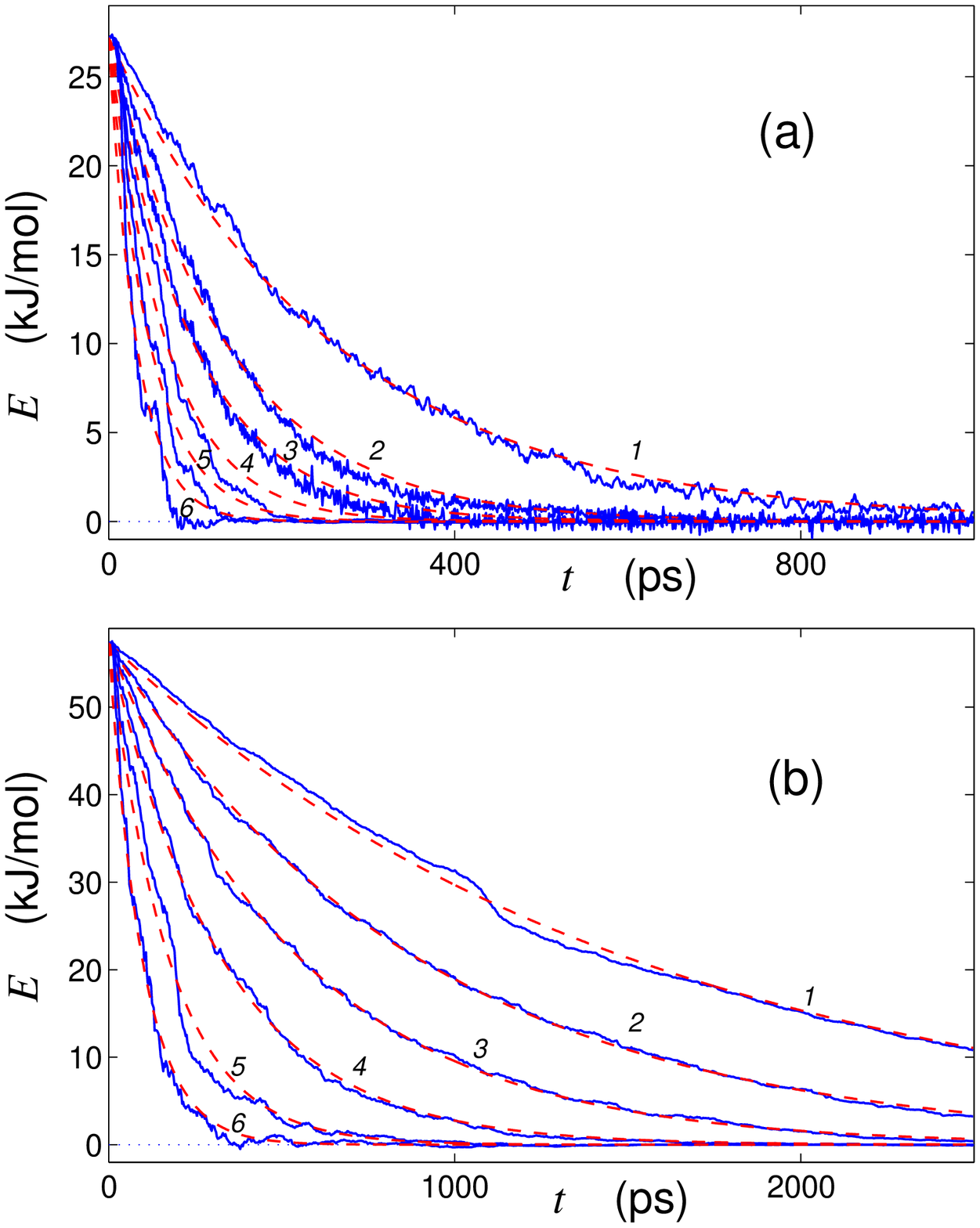}
\end{center}
\caption{\label{fg06}\protect
        Decreasing the breather energy $E$ in thermalized chain with
        $\Lambda=130.1$ kJ/mol, $\omega=820.5$ cm$^{-1}$ (a) and
        $\Lambda=529$ kJ/mol, $\omega=928$ cm$^{-1}$ (b) under temperature
        $T=1$, 2, 3, 5, 10, 20 K (curves 1, 2,...,6).
        Solid lines determine temporal dependencies of breather energy for concrete
        realization of chain thermalization, dotted lines -- corresponding exponential
        law $E(t)=E(0)\exp(-\alpha t)$.
        }
\end{figure}

To estimate the breather time of life in thermalized chain let consider the
temporal dependence of the energy (Fig. \ref{fg06} and \ref{fg10}). It is seen,
that the energy decreases monotonously by exponential low
$E(t)=E(0)\exp(-\alpha t)$. One can determine the breather time of life as
half of breaking period $t_\alpha=\ln 2/\alpha$b $(E(t_\alpha)=E(0)/2)$.
Dependence of $t_\alpha$ on the chain temperature $T$ is presented in
table \ref{tb1}.
We conclude, that breather time of life is proportional to ratio of its
energy to temperature.

%-------------------------------------------------------------------------
\begin{table}[tb]
\caption{\label{tb1}
Dependence of breathers time of life $t_\alpha$ on the chain temperature $T$
(frequency $\omega=820.5$ cm$^{-1}$, parameter $\Lambda=130.122$ kJ/mol and
$\omega=928.4$ cm$^{-1}$, $\Lambda=529$ kJ/mol).}
\begin{center}
\begin{tabular}{cccccccc}
\hline\hline
$\Lambda$ (kJ/mol) & $T$ (K) &  1 & 2 & 3 & 5 & 10 & 20  \\
\hline
130.122 & $t_\alpha$ (ps) &  180 & 94 & 68 & 46 & 33 & 22 \\
529     & $t_\alpha$ (ps) & 1052 & 625 & 386 & 228 & 121 & 68 \\
\hline\hline
\end{tabular}
\end{center}
\end{table}
%-------------------------------------------------------------------------

%---------------------------- Fig. 6 ------------------------------------
%---------------------------- Fig. 6 ------------------------------------

Besides the considered breathers, the supersonic acoustic solitons can exist in
PE chain [17]. Therefore, it is desirable to study their interaction with optic
breathers [we choose further for definiteness the system of parameters
(\ref{n2})]. In such a case the acoustic solitons are caused by a local
longitudinal compression of the chain (the properties of acoustic soliton were
presented in  \cite{Manevitch97}).

\begin{figure}[tb]
\begin{center}
\includegraphics[angle=0, width=.8\linewidth]{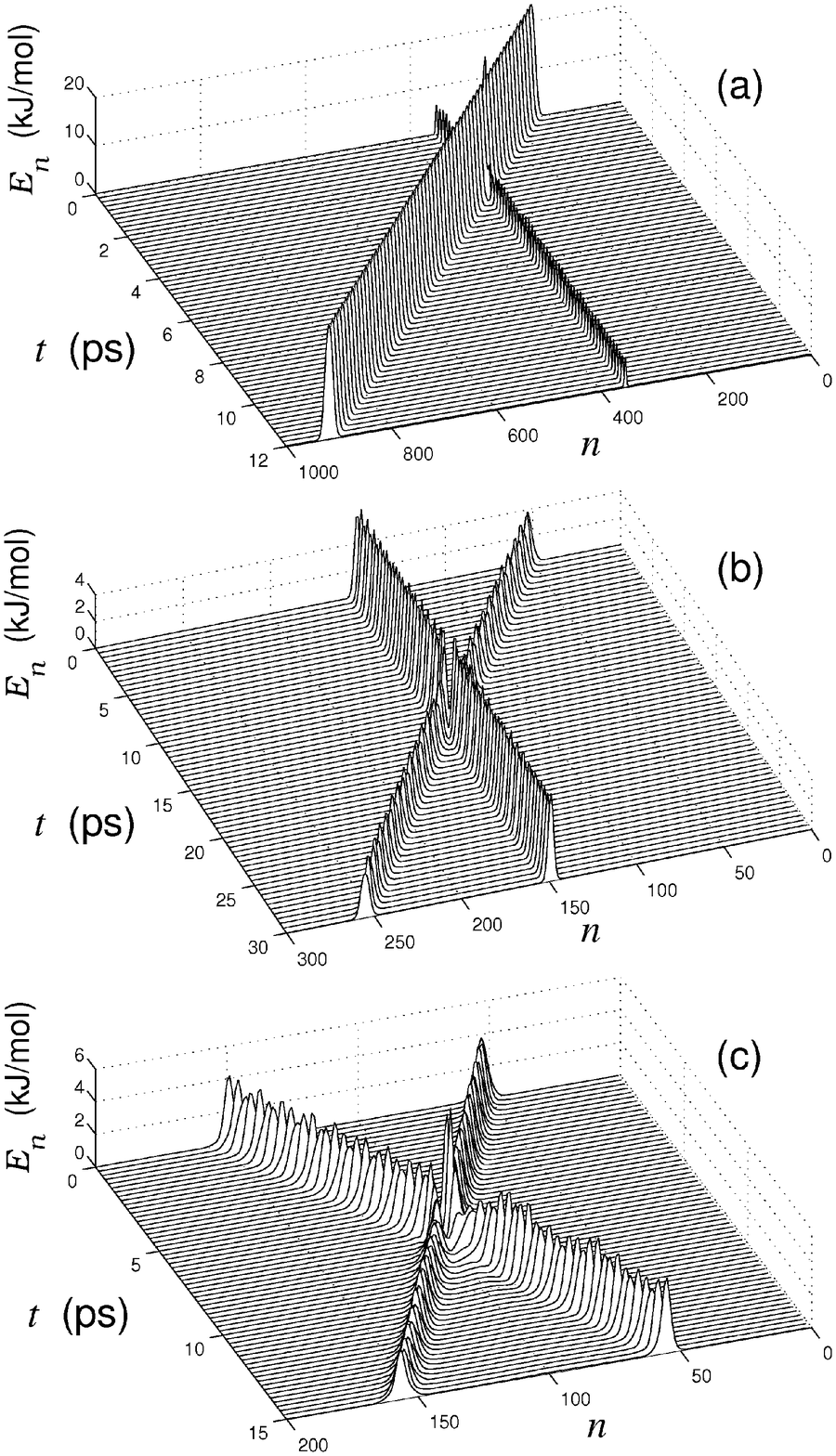}
\end{center}
\caption{\label{fg07}\protect
        Collision of supersonic acoustic soliton with stationary optic breather
        ($\omega=820.5$ cm$^-1$) (a), collision of moving breather (velocity
        $s=0.106s_0$) with stationary one (b) and elastic collision of two
        moving breathers (c). The temporal dependence of energy distribution
        $E_n$ along the chain is shown.
        }
\end{figure}

The collision of acoustic soliton with the stationary breather is shown at
Fig. \ref{fg07} (a). It is seen, that considered interaction does not lead
to noticeable change of soliton energy, but the breather acquires a small momentum
in the direction of the soliton motion. E.g., after collision of the soliton
with velocity $s=1.024s_0$, where $s_0=7790$ m/s -- velocity of long wavelength
optic phonons, the breather with frequency $\omega=820.5$ cm$^{-1}$ becomes to move
with constant velocity $s=0.0255s_0$. After every new collision, the breather velocity
increases attaining the value $s=0.106s_0$ in limit.
%---------------------------- Fig. 7 ------------------------------------
%---------------------------- Fig. 7 ------------------------------------

\begin{figure}[tb]
\begin{center}
\includegraphics[angle=0, width=1\linewidth]{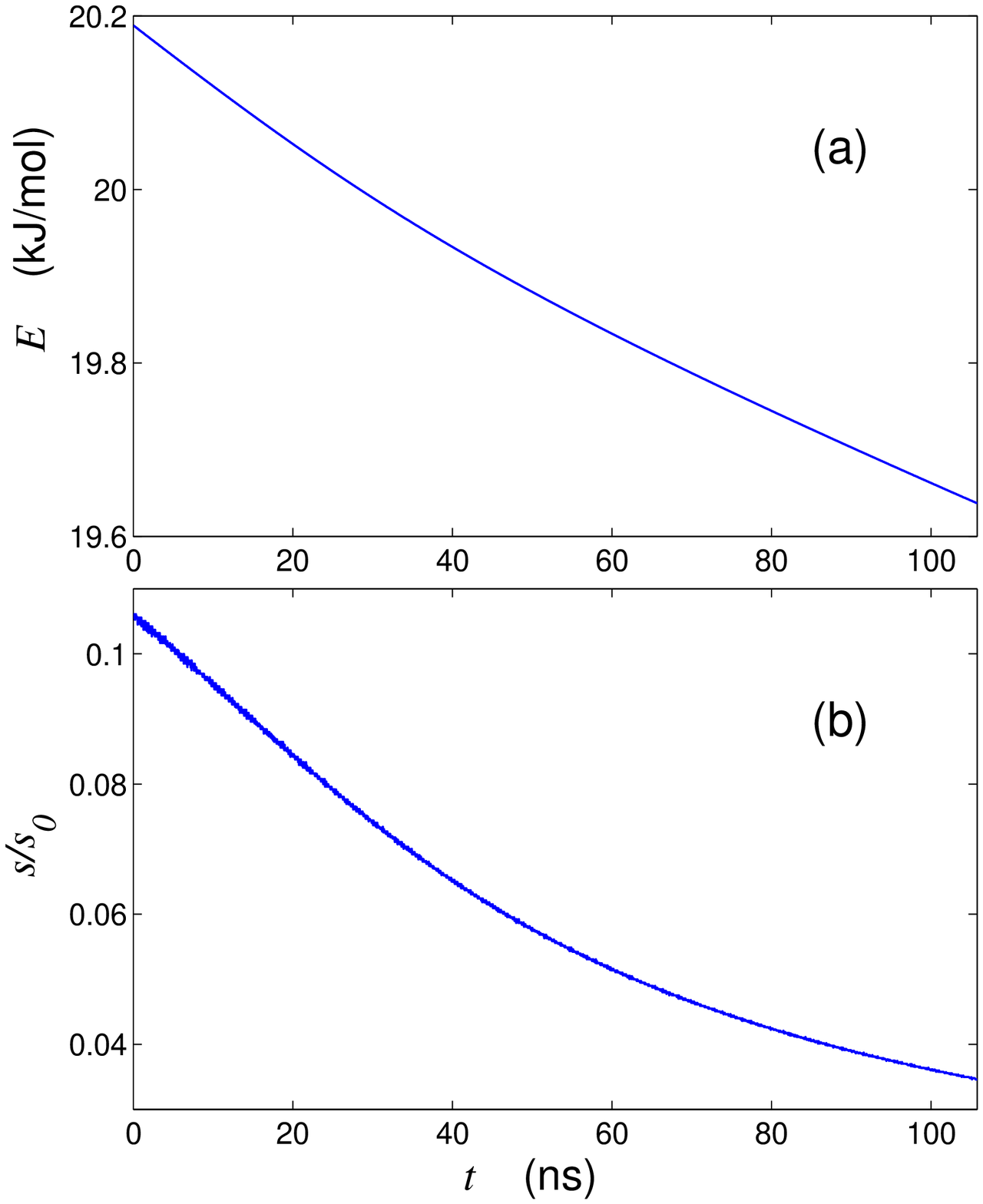}
\end{center}
\caption{\label{fg08}\protect
        Decreasing the breather energy $E$ (a) and dimensionless breather
        velocity $s/s_0$ (b) in PE chain.
        }
\end{figure}

The motion of the breather is accompanied by phonon irradiation with low
intensity that leads to decreasing of breather energy. The temporal
dependence of the breather energy is shown at \ref{fg08}. It is well seen that
change of velocity and energy decreasing may be fixed at very large time only ($\sim
10$ ns).
%---------------------------- Fig. 8 ------------------------------------
%---------------------------- Fig. 8 ------------------------------------

It is necessary to note that propagating breather has the
frequency slightly exceeding the low boundary of frequency
spectrum. So, we deal here with the breathers in the propagation
zone of optic spectrum. However, the time of life in this case
turns out also to be large enough as well as in the case of
breather in the gap between acoustic and optic branches of the
dispersion curve. Actually, let us consider the collision of
propagating breather ($\omega=830$ cm$^{-1}$) with stationary one
($\omega=820$ cm$^{-1}$) [see Fig. \ref{fg07} (b)] and collision
of two propagating breathers ($\omega=830$ cm$^{-1}$) [see Fig.
\ref{fg07} (c)]. One can see that the breathers with frequency in
propagation zone can move freely without any noticeable changes
and interact with similar or stationary ones as elastic particles
-- they exchange by momentum without any energy decreasing.

\begin{figure}[tb]
\begin{center}
\includegraphics[angle=0, width=1\linewidth]{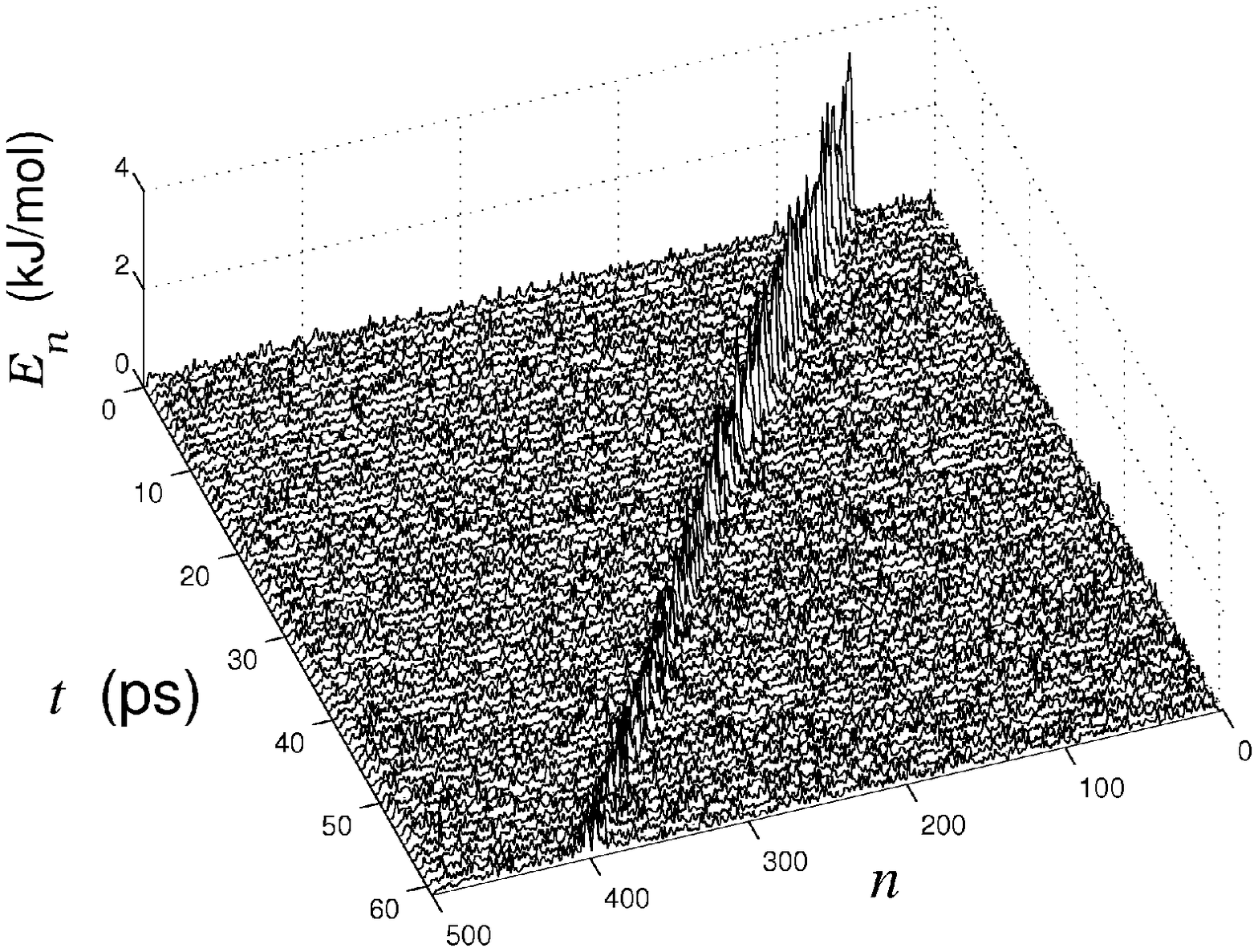}
\end{center}
\caption{\label{fg09}
        Breaking of moving breather (velocity $s=0.106s_0$) in thermalized
        chain (temperature $T=5$K). Temporal dependence of energy distribution
        along the chain $E_n$ is shown.
        }
\end{figure}

\begin{figure}[tb]
\begin{center}
\includegraphics[angle=0, width=.7\linewidth]{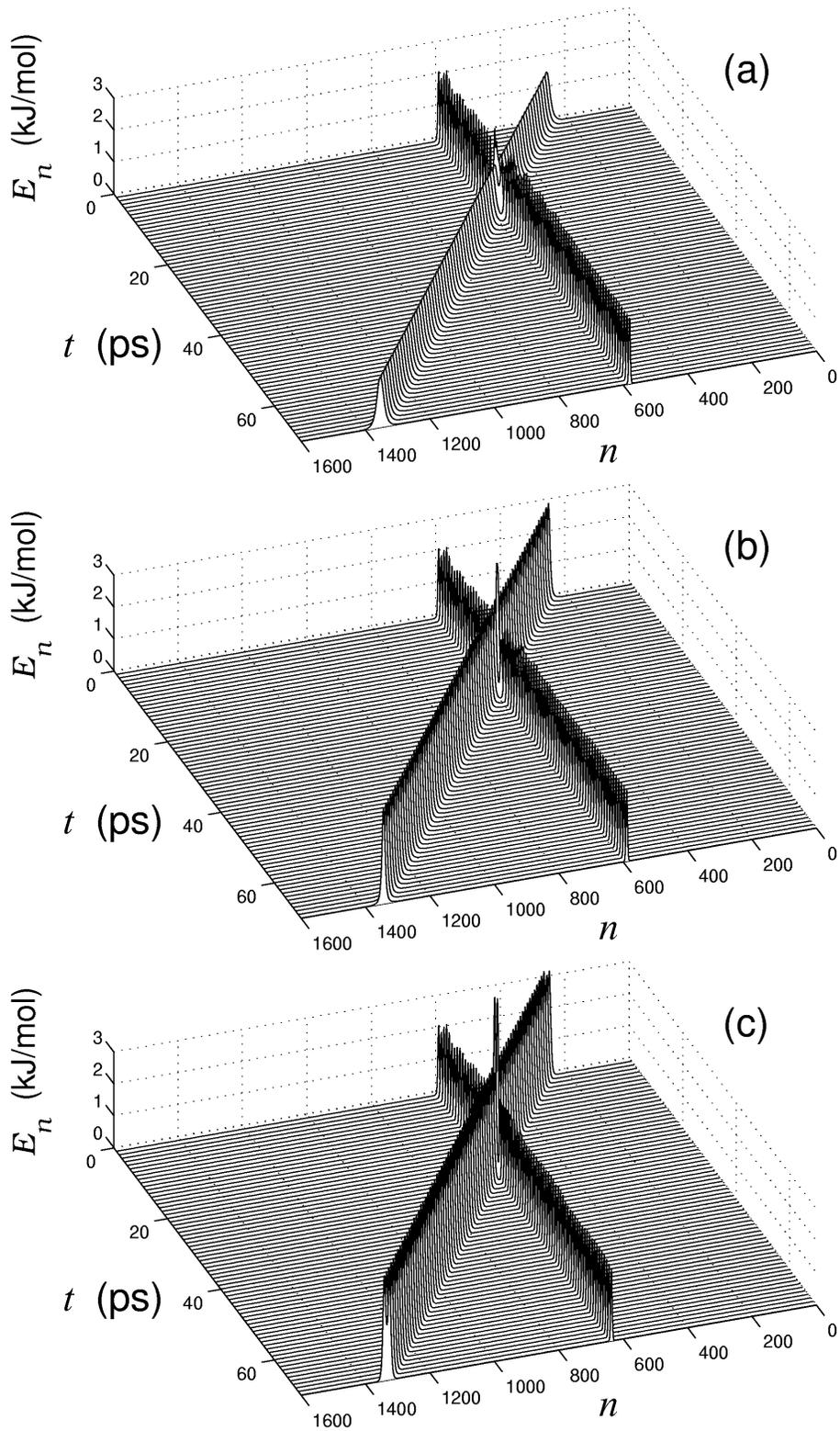}
\end{center}
\caption{\label{fg10}
        Passage of topological solitons (velocity $s = 0.25s_0$) with charge
        ${\rm\bf q}=(1, 0)$ (a), ${\rm\bf q}=(0.5, 0.5)$ (b),
        and ${\rm\bf q}=(0, 1)$ (c) through static breather
        (frequency $\omega=820.5$ cm$^{-1}$).}
\end{figure}

In thermalized chain both propagating and stationary breathers
have the time of life, which is proportional to ratio of breathers
energy to temperature and does not noticeably depend on the
breather velocity (Fig. \ref{fg09}).
%---------------------------- Fig. 9 ------------------------------------
%---------------------------- Fig. 9 ------------------------------------

We have shown that the optic breathers can exist in both isolated chain and
chain interacting with neighbor ones in the crystals. As this takes place the
characteristics of the breather are not depend noticeably on the intermolecular
interactions. In this case we can use approximation of immovable neighbor
chains. Corresponding Hamiltonian
(\ref{f3}) has a view
\begin{eqnarray}
H &=&\sum_{n=-\infty}^{+\infty}\{\frac12m [\dot u_n^2 + \dot v_n^2 + \dot w_n^2]
\nonumber\\
&&+[p_1 + p_2 \cos(\phi_n) + p_3\cos(3\phi_n)]\label{n5}\\
&&+\frac12\Lambda[\cos(\theta_n)-\cos(\theta_0)]^2\nonumber \\
&& + {\cal D}[1 - e^{-q(\rho_n - \rho_0)}]^2+W(u_n,v_n,w_n)\},
\nonumber
\end{eqnarray}
where the function $W(u_n,v_n,w_n)$ describes the interchain interaction.
Detail description of this approach is given in  \cite{p21}.
%---------------------------- Fig. 10 ------------------------------------
%---------------------------- Fig. 10 ------------------------------------
%---------------------------- Fig. 11 ------------------------------------
\begin{figure}[tb]
\begin{center}
\includegraphics[angle=0, width=1\linewidth]{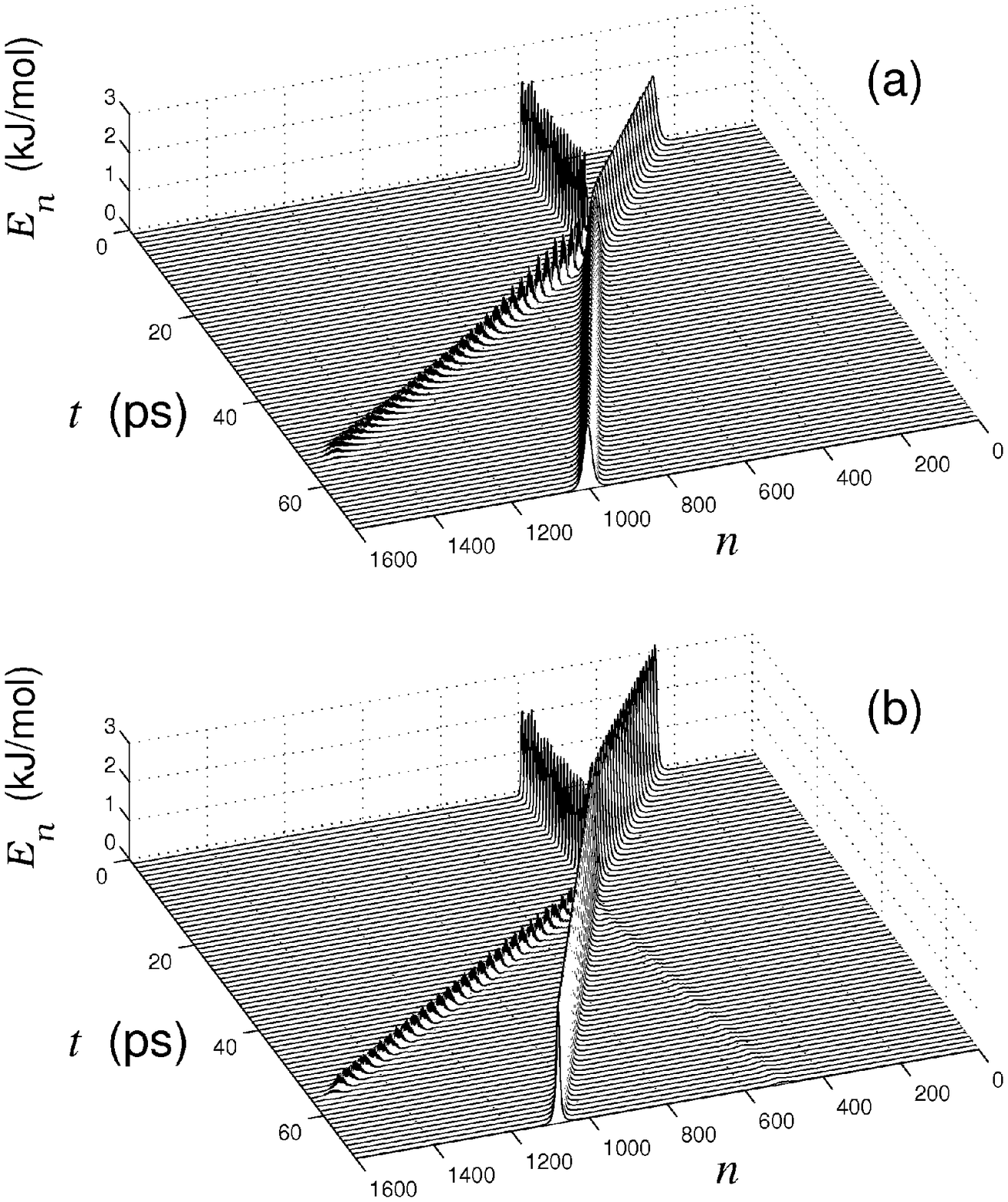}
\end{center}
\caption{\label{fg11}
        Breaking of the stationary breather (frequency $\omega=820.5$ cm$^{-1}$)
        as a result of collision with topological solitons with integer charge
        ${\rm\bf q}=(-1, 0)$ (a) and half-integer charge ${\rm\bf q}=(-0.5, 0.5)$
        (b) (velocity of topological soliton $s = 0.25s_0$).}
\end{figure}
%---------------------------- Fig. 11 ------------------------------------
%---------------------------- Fig. 12 ------------------------------------
\begin{figure}[tb]
\begin{center}
\includegraphics[angle=0, width=0.8\linewidth]{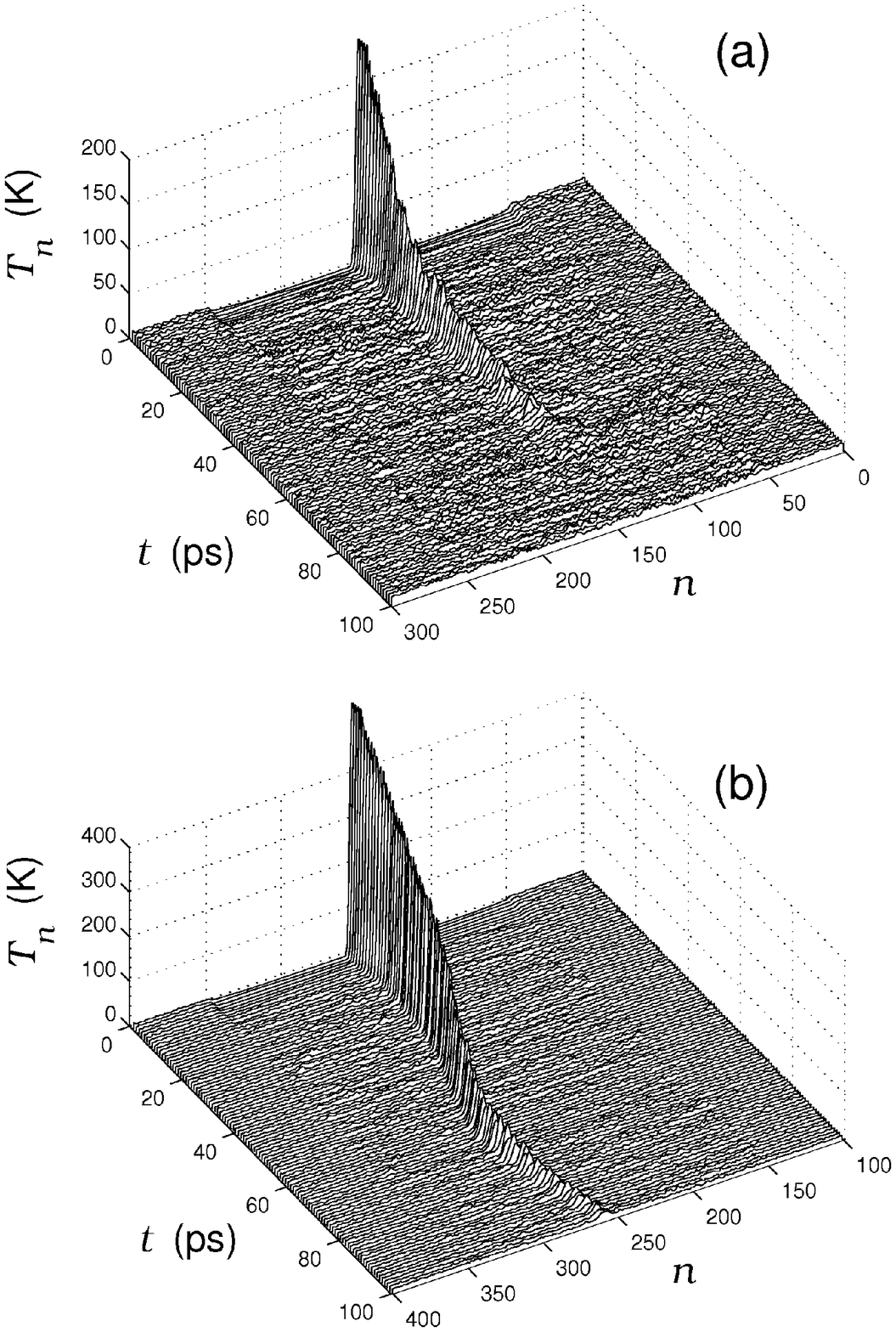}
\end{center}
\caption{\label{fg12}
        Breaking of the stationary (a) and bound state (b) of the stationary
        breather with topological soliton [charge ${\rm\bf q}=(1, 0)$] in
        thermalized chain with T=10K.}
\end{figure}
%---------------------------- Fig. 12 ------------------------------------

If taking into account interchain interaction, three types of topological
solitons with topological charges ${\bf q}=(q_1,q_2)$ appear. Their properties
are described
in  \cite{p21}. The solutions of the first type have topological charge
{\bf q}$=(\pm 1,0)$ and describe a localized longitudinal extension
(compression) of the chain by one period. The solutions of the second type
have the charge {\bf q}$=(\pm 0.5,0.5)$ and describe a longitudinal extension
(compression) of the chain by halved period and simulations twist by
$180^\circ$. The solutions of the third type have the charge {\bf q}$=(0,1)$
and describe the twist of the chain by  $360^\circ$.
All these topological solutions have subsonic spectra of velocities and
can move without phonon irradiation.

We considered an interaction of stationary optic breather having
frequency $\omega=820.5$ cm$^{-1}$ with moving topological
solitons. As it is seen from Fig. \ref{fg10}  such an interaction
is elastic one if first component of topological charge $q_1\le
0$, i.e. if a compression in the localization region is absent. In
opposite case, when $q_1<0$, the interaction with topological
soliton leads to breaking the breather -- see Fig. \ref{fg11}.
Moreover, the bound state of breather and topological soliton
($q_1>0$) may be energetically profitable. Such a coupling
increases the lifetime of breather in thermalized chain [compare
Fig. \ref{fg12} (a) and (b)].

\section{Conclusion}
The optic breathers which are localized coupled
longitudinal-transversal nonlinear excitations can exist in both attenuation
and propagation zones of PE
crystal. In spite approximate analytical solution for breathers
has been obtained using a model of isolated chain, we confirmed
numerically not only validity of such a model but revealed also
that the parameters of breathers change unnoticeably if taking
into account interchain interaction. The reason is a
weakness of interchain interaction in comparison to intrachain
one. Such a weakness is especially clear for optic excitations.
The breathers demonstrated stability
to mutual collision and have large enough time of life in the presence
of thermal excitations. The interaction of optic breathers with
supersonic and subsonic (topological) solitons may be both elastic
and inelastic dependent on their parameters.
%\bibliography{ref1}

The authors (L.~I.~M. and A.~V.~S.) thank the Region Rhone-Alpes and
the Russian Foundation of Basic Research (awards
04-02-17306 and 04-03-32119), respectively for financial support.

\end{document}